\documentclass[twocolumn,aps,prl,amsmath,amssymb]{revtex4-2}
\usepackage{graphicx}
\usepackage{dcolumn}
\usepackage{bm}

\usepackage{multirow}

\usepackage{amssymb}
\usepackage{amsmath}
\usepackage{graphicx}
\usepackage{xcolor}
\usepackage[colorlinks,citecolor=blue,linkcolor=red,urlcolor=blue]{hyperref}
\usepackage{braket}
\usepackage{CJK}
\usepackage{indentfirst}
\usepackage{amsmath}
\usepackage{cases}
\usepackage{enumerate}
\usepackage{booktabs} 
\usepackage{tabularx} 

\def\Eq#1{Eq.~(\ref{#1})}

\begin{document}
\title{Finite-time scaling beyond the Kibble-Zurek prerequisite in Dirac systems}
\author{Zhi Zeng$^{1,2}$\footnotemark[1]}
\altaffiliation{These authors contribute equally to this work.}
\author{Yin-Kai Yu$^{1,2,3,4}$\footnotemark[1]}
\altaffiliation{These authors contribute equally to this work.}
\author{Zhi-Xuan Li$^{1,2}$\footnotemark[1]}
\altaffiliation{These authors contribute equally to this work.}
\author{Zi-Xiang Li$^{3,4}$\footnotemark[1]}
\email{zixiangli@iphy.ac.cn}
\author{Shuai Yin$^{1,2}$}
\email{yinsh6@mail.sysu.edu.cn}

\affiliation{$^1$Guangdong Provincial Key Laboratory of Magnetoelectric Physics and Devices, School of Physics, Sun Yat-Sen University, Guangzhou 510275, China}
\affiliation{$^2$School of Physics, Sun Yat-Sen University, Guangzhou 510275, China}
\affiliation{$^3$Beijing National Laboratory for Condensed Matter Physics \& Institute of Physics, Chinese Academy of Sciences, Beijing 100190, China}
\affiliation{$^4$University of Chinese Academy of Sciences, Beijing 100049, China}

\date{\today}
\begin{abstract}
The conventional Kibble-Zurek mechanism and the finite-time scaling provide universal descriptions of the driven critical dynamics from gapped initial states based on the adiabatic-impulse scenario. Here we investigate the driven critical dynamics in two-dimensional Dirac systems, which harbor semimetal and Mott insulator phases separated by the quantum critical point triggered by the interplay between fluctuations of gapless Dirac fermions and order parameter bosons. We find that despite the existence of the gapless initial phase, the driven dynamics can still be captured by the finite-time scaling form. This leads us to propose a criterion for the validity of Kibble-Zurek mechanism with a gapless initial state. Accordingly, our results generalize the Kibble-Zurek theory to incorporate composite fluctuations and relax its requirement for a gapped initial state to systems accommodating gapless Dirac fermionic excitations. Our work not only brings fundamental perspective into the nonequilibrium critical dynamics, but also provides an approach to fathom quantum critical properties in fermionic systems.
\end{abstract}
\footnotetext[1]{These authors contribute equally to this work.}
\maketitle

\noindent {\bf INTRODUCTION}

Fathoming nonequilibrium universal properties near a quantum critical point (QCP) is one of the central issues in modern physics~\cite{Dziarmaga2010review,Polkovnikov2011rmp}. Although the general organizing principle for the nonequilibrum critical dynamics is still elusive, a unified framework for understanding the generation of topological defects after the linear quench was proposed by Kibble in cosmological physics and then generalized by Zurek in condensed matter systems ~\cite{Kibble1976,Zurek1985}. This celebrated Kibble-Zurek mechanism (KZM) has aroused intensive investigations from both theoretical and experimental aspects, exerting far-reaching significance in both classical and quantum phase transitions~\cite{Kibble1976,Zurek1985,Zoller2005prl,Dziarmaga2005prl,PhysRevB.72.161201,Du2023,Ko2019,PhysRevLett.129.227001,Keesling2019,Ebadi2021,sciadv.aba7292,science.abo6587,science.abq6753,PRXQuantum}. More interestingly, it was found that scaling behaviors can also manifest themselves in the driven process~\cite{Zhifangxu2005prb,Deng2008epl,Chandran2012prb,Clark2016science,Huse2012prl}. As a generalization of the KZM, a finite-time scaling (FTS) theory was proposed to systematically understand the full scaling properties~\cite{Gong2010njp,Feng2016prb}. These full scaling forms have been verified in various systems from numerical to experimental works~\cite{huangyy2014prb,Liuchengwei2014prb,Yin2014prb,Sandvik2015prl,Clark2016science,Gong2010njp,Feng2016prb,king2023nature,Keesling2019,Ebadi2021,garcia2024resolving}. Moreover, the KZM and the FTS recently show their fabulous power in state preparations and probing critical properties in fast-developing programmable quantum devices~\cite{king2023nature,Keesling2019,Ebadi2021,garcia2024resolving,PhysRevB.106.L041109}.

At the core of the original KZM lies the adiabatic-impulse scenario~\cite{Dziarmaga2010review,Polkovnikov2011rmp,Polkovnikov2008natphy}. According to it, a crucial prerequisite for the implementation of the KZM is the existence of a gapped initial stage, wherein the system evolves adiabatically along the equilibrium state. The border of this initial stage with the intrinsically nonequilibrium impulse region gives rise to a frozen time, which dominates the critical dynamics near the critical point, yielding the FTS forms~\cite{Kibble1976,Zurek1985,Polkovnikov2011rmp,Dziarmaga2010review,Gong2010njp,Feng2016prb}. Moreover, intriguing dynamic scaling behavior dominated by the QCP was also found for driven dynamics from gapless initial phase to cross the QCP in one dimensional spin systems~\cite{PhysRevB.78.144301,PhysRevB.92.064419}. However, the universal criterion for the validity of KZM and FTS with a gapless initial state has yet to be established.

The QCP occurring in strongly interacting Dirac systems, dubbed as Dirac QCP, represents a typical class of QCP which has joint critical fluctuations from both order parameter and gapless fermions. Studies of Dirac QCP stem from the research in modern high-energy physics, such as chiral symmetry in QCD and mass generation via spontaneous symmetry breaking~\cite{Gross1974prd,Gracey2016prd,Poland2019rmp,Youyz2018prx}. From the perspectives of statistical mechanics and condensed matter physics, the Dirac QCP also attracts enormous attentions, particularly after the experimental realization of two-dimensional Dirac fermions in graphene and various topological insulators or semimetals~\cite{Geim2009rmp,KaneReview,SCZhangReview}. The presence of Dirac fermions is theoretically revealed to tremendously enrich quantum critical properties, rendering novel universality classes of quantum phase transition without classical counterpart~\cite{Janssen2014prb,Parisen2015prb,Li2018ScienceAdvances,Vaezi2022PRL,Sorella2018PRB,Sorella1992,Herbut2006prl,Herbut2013prx,PhysRevB.91.165108,Sorella2016prx,Wang2014NJP,Li2015NJP,Yao2017nc}. 

Exotic aspects of Dirac QCP in equilibrium also suggest exotic phenomena out of equilibrium. Interesting dynamic scaling behaviors were discovered in QCPs that feature non-interacting Dirac fermions~\cite{dutta_quenching_2010,PhysRevB.106.134203,PhysRevLett.134.010409}. However, nonequilibrium dynamics in strongly interacting Dirac QCP is still largely unexplored. Particularly, for driven dynamics along the gapless Dirac semimetal (DSM) phase to cross the QCP, whether the original KZM is still applicable remains unknown. Consequently, investigating the nonequilibrium driven dynamics in Dirac QCP has overarching meaning in fundamental theory, as well as immediate applications in the context of detecting and exploring fermionic QCP in experimental platforms~\cite{Geim2009rmp,KaneReview,SCZhangReview}.

However, directly tackling the real-time dynamics in two or higher spatial dimension is largely hindered by the lack of reliable theoretical or numerical methods. Specifically, quantum Monte Carlo (QMC) fails as a result of the notorious sign problem~\cite{AssaadReview,Li2019Review}, while the tensor-network method still needs tremendous improvements despite remarkable progress in recent years~\cite{Dziarmaga2022sciadv}. Fortunately, scaling analysis demonstrates that both real- and imaginary-time driven dynamics share the same scaling form~\cite{Polkovnikov2011prb} (See Supplementary Materials, Sec. I). This inference has been verified in various systems~\cite{Sandvik2015prl,Polkovnikov2011prb,king2023nature}, bridging the gap between the QMC imaginary-time simulations without sign problem and the real-time dynamics (See Supplementary Materials, Sec. II).

In this work, for the first time we investigate the driven critical dynamics of two representative strongly interacting Dirac QCPs, belonging to chiral Heisenberg and chiral Ising universality classes, respectively, via the determinant QMC method~\cite{AssaadReview}. By linearly varying the interaction strength along the imaginary-time direction to cross the QCP from both the DSM and Mott insulator phases, we uncover that the driven process near the Dirac QCP satisfies the scaling form of FTS despite the violation of the adiabatic-impulse scenario of the KZM due to the existence of the gapless initial state. Furthermore, we develop a generalized criterion for the validity of the KZM and FTS with the gapless initial state. In addition, our numerical simulation achieves the critical exponents of the Dirac QCP, whose values obtained in previous studies on equilibrium properties are still under debate. Through the innovative generalization of quantum driven critical dynamics to strongly interacting Dirac QCP, our study not only leads to a great leap to the fundamental theory of KZM and FTS, but also contributes a feasible approach to investigate the fermionic quantum critical phenomena in realistic platform such as quantum materials and devices.

\noindent {\bf RESULTS}

\noindent {\bf Dynamics in chiral Heisenberg criticality}

A typical model hosting Dirac QCP belonging to chiral Heisenberg universality class is the Hubbard model on the half-filled honeycomb lattice with the Hamiltonian~\cite{Sorella1992,Herbut2006prl,Herbut2013prx,PhysRevB.91.165108,Sorella2016prx}
\begin{equation}
 H=-t\sum_{\langle ij\rangle,\sigma}c_{i\sigma}^\dagger c_{j\sigma}+U\sum_i \left({n_{i\uparrow}-\frac{1}{2}}\right) \left({n_{i\downarrow}-\frac{1}{2}}\right) \label{eq:Hamiltonian},
\end{equation}
in which $c_{i\sigma}^\dagger$ ($ c_{j\sigma}$) represents the creation (annihilation) operator of electrons with spin $\sigma$, $n_{i\sigma}\equiv c_{i\sigma}^\dagger c_{i\sigma}$ is the electron number operator, $t$ is the hopping amplitude between nearest neighbor sites and set as the energy unit in the following, and $U$ is the strength of the on-site repulsive interaction. The model is absent from sign problem in QMC simulation (See Supplementary Materials, Sec. II). As shown in Fig.~\ref{figure1}, a critical point $U_c\approx 3.85$ separates two phases~\cite{Sorella2016prx}. When $U>U_c$, the system is in the antiferromagnetic (AFM) Mott insulator phase in which fermions acquire a mass originating from spontaneous symmetry breaking characterized by the finite AFM order parameter $m^2 = \sum_{i,j}\eta_i \eta_j \langle{S_i^z S_j^z}\rangle/L^{2d}$ with $S_i^{z}\equiv\frac{1}{2}\bm{c}^\dagger_{i}\sigma^z\bm{c}_{i}$, ${\bm{c}\equiv({c_\uparrow,c_\downarrow}})$, and $\eta_i=\pm 1$ for $i \in A(B)$ sublattice~\cite{Herbut2013prx,PhysRevB.91.165108,Sorella2016prx}. Here, $L$ is the linear size of the system and $d=2$ is the spatial dimension. In this phase, the transverse spin excitation is massless due to the presence of the Goldstone modes. In contrast, when $U<U_c$, the system is in the DSM phase with four-component massless Dirac fermion ($N_f=2$). At $U_c$, both Dirac fermions and AFM order parameter bosons are gapless, yielding the Gross-Neveu QCP belonging to chiral Heisenberg universality class~\cite{Sorella1992,Herbut2006prl,Herbut2013prx,PhysRevB.91.165108,Sorella2016prx}.

\begin{figure}[tbp]
\centering
  \includegraphics[width=\linewidth,clip]{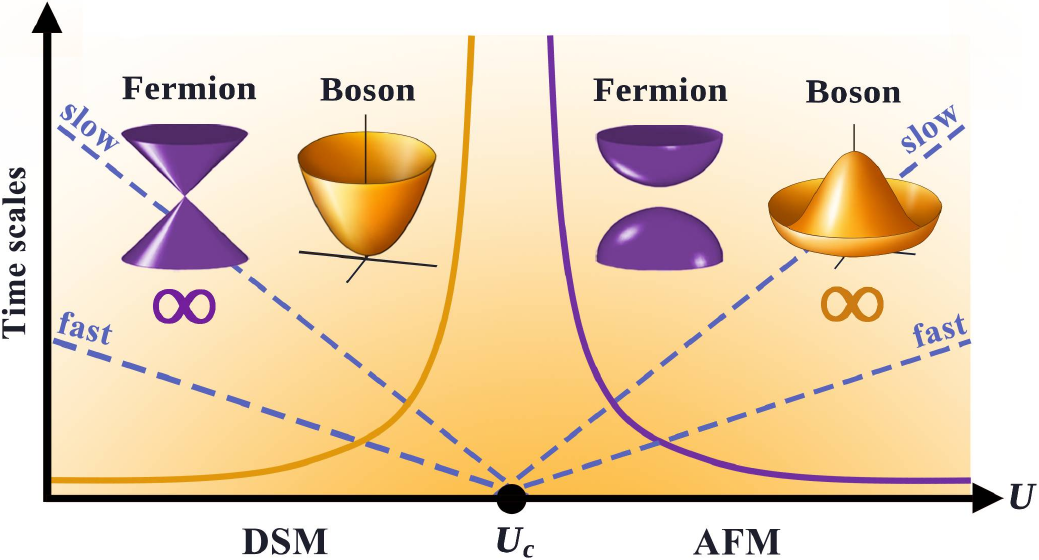}
  \vskip-3mm
  \caption{Sketch of the phase diagram in Dirac systems and the protocol for driven dynamics with different initial states. The correlation time scales for both boson (yellow solid curve) and fermion (violet solid curve) are finite in one phase but divergent (symbolized by ``$\infty$'') in the other phase. The dashed line denotes the time distance to the critical point for different driving rates. Accordingly, the prerequisite of the original KZM that a gapped initial state should exist to protect an initial adiabatic stage, in which the transition time is larger than the correlation time, breaks down.
  }
  \label{figure1}
\end{figure}

\begin{figure}[tbp]
\centering
  \includegraphics[width=\linewidth,clip]{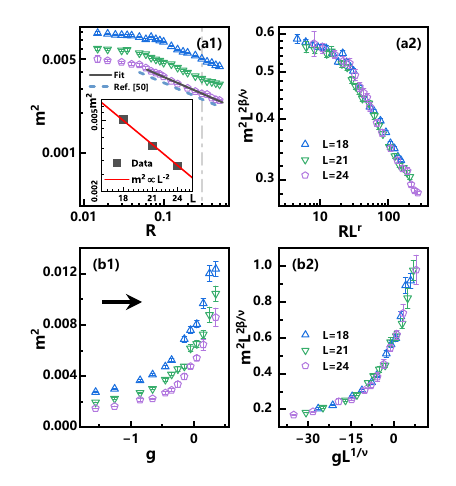}
  \vskip-3mm
  \caption{Driven dynamics from the DSM phase of model~(\ref{eq:Hamiltonian}). \textbf{(a)} Log-log plots of $m^2$ versus $R$ for different $L$ driven to $U_c=3.85$ before (a1) and after (a2) rescaling. Inset in (a1) shows $m^2 \propto L^{-2}$ for $R=0.3$ (dash-dotted line). For large $R$, power fitting for $L=24$ (black solid line) shows $m^2 \propto R^{-0.26(1)}$ with the exponent close to $(2\beta-d \nu)/\nu r=-0.26(2)$ (dash line) from Ref.~\cite{Sorella2016prx}. \textbf{(b)} Curves of $m^2$ versus $g$ for fixed $RL^r=5.41$ and different $L$ before (b1) and after (b2) rescaling. The arrow indicates the driving direction.
  }
  \label{figure2}
\end{figure}

Here we begin to explore whether the FTS forms are still applicable in the driven dynamics of this Dirac QCP with composite critical fluctuations from gapless initial states. First we study the driven dynamics by varying $U$ with imaginary time $\tau$ as $U=U_0+R \tau$ from the DSM initial state with $U_0=0$, as illustrated in Fig.~\ref{figure1}. We denote the distance to the critical point as $g$ (here $g=U-U_c$). The smallest lattice size is chosen as $L=18$ to eliminate the scaling violation induced by small sizes. When $g=0$, from Fig.~\ref{figure2} (a1), we find that for large $R$, $m^2\propto L^{-2}R^{-0.26(1)}$ with the exponent on $R$ close to $(2\beta-d\nu)/\nu r=-0.26(2)$, in which $\beta=0.76(2)$ and $\nu=1.02(1)$ are the exponents for order parameter and correlation length, respectively~\cite{Sorella2016prx}, and $r=z+1/\nu$ is the scaling dimension of $R$. Here the dynamic exponent $z$ equals one in the Gross-Neveu Dirac QCP owing to the Lorentz symmetry of the effective model~\cite{Herbut2006prl}. Additionally, the exponent on $R$ is almost independent of $L$. In contrast, when $R$ is small,  Fig.~\ref{figure2} (a1) shows that $m^2$ tends to saturate and the usual finite-size scaling $m^2\propto L^{-2\beta/\nu}$ is restored. To reconcile these rescaling relations, the scaling form must satisfy
\begin{equation}
  m^2(R,L,g)=L^{-d} R^{(2\beta-d\nu)/\nu r} \mathcal{F}(RL^{r},gL^{1/\nu}),
  \label{eq:scaling1}
\end{equation}
in which $\mathcal{F}$ is a non-singular scaling function and therein dimensionless quantity $gL^{1/\nu}$ is also included to take account of the off-critical-point effects. Eq.~(\ref{eq:scaling1}) is consistent with the FTS in conventional bosonic QCP~\cite{huangyy2014prb,Liuchengwei2014prb}.

To confirm Eq.~(\ref{eq:scaling1}), we find that \Eq{eq:scaling1} yields the scaling form $m^2(R,L)=L^{-2\beta/\nu} \mathcal{F}_1(RL^{r})$ at $g=0$. Here, we rescale $m^2$ and $R$ as $m^2L^{2\beta/\nu}$ and $RL^r$ with the exponents $\beta=0.76(2)$, $\nu=1.02(1)$~\cite{Sorella2016prx}, and $z=1$~\cite{Herbut2006prl} set as input, and reveal that the rescaled curves collapse well into a single curve, as shown in Fig.~\ref{figure2} (a2), confirming Eq.~(\ref{eq:scaling1}) at $g=0$. 

\begin{figure}[tbp]
\centering
  \includegraphics[width=\linewidth,clip]{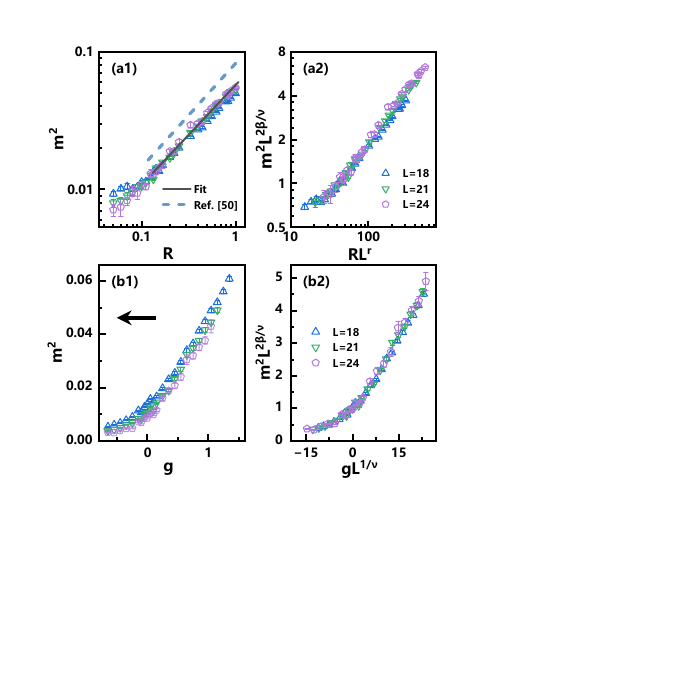}
  \vskip-3mm
  \caption{Driven dynamics from the AFM phase of model~(\ref{eq:Hamiltonian}). \textbf{(a)} Log-log plots of $m^2$ versus $R$ for different $L$ driven to $U_c=3.85$ before (a1) and after (a2) rescaling. For large $R$, power fitting (black solid line) shows $m^2 \propto R^{0.73(2)}$ with the exponent close to $2\beta/\nu r=0.75(2)$ (dash line) from Ref.~\cite{Sorella2016prx}. \textbf{(b)} Curves of $m^2$ versus $g$ for fixed $RL^r=41.5$ and different $L$ before (b1) and after (b2) rescaling. The arrow indicates the driving direction.
  }
  \label{figure3}
\end{figure}

In addition, to unravel the scaling properties in the driven process near $U_c$, we calculate the dependence of $m^2$ on $g$ for an arbitrary fixed $RL^r=5.41$ and present the results in Fig.~\ref{figure2} (b1). After rescaling $m^2$ and $g$ by $L^{2\beta/\nu}$ and $L^{1/\nu}$, the curves with various $L$ collapse into each other, as displayed in Fig.~\ref{figure2} (b2), confirming that the universal scaling behavior of physical observable in the driven process is described by Eq.~(\ref{eq:scaling1}).

The reason for the appearance of the scaling relation $m^2\propto L^{-d} R^{(2\beta-d\nu)/\nu r}$ is that for large $R$, driven induced length scale $\xi_R \sim R^{-1/r}$ is smaller than $L$. Thus, the definition of $m^2$ indicates that $m^2\propto L^{-d}$ owing to the central limit theorem. Meanwhile, the rest part of the dimension of $m^2$ should be borne by $R$, giving rise to the leading term of Eq.~(\ref{eq:scaling1}). In this case, $\mathcal{F}(RL^{r},0)$ tends to a constant. In contrast, for small $R$, $\xi_R>L$, such that the conventional finite-size scaling at equilibrium $m^2\propto L^{-2 \beta/\nu}$ is recovered, and $\mathcal{F}(RL^r,0)$ obeys $\mathcal{F}(RL^{r},0)\sim (RL^{r})^{d/r - 2\beta/\nu r}$.

\begin{figure}[tbp]
\centering
  \includegraphics[width=\linewidth,clip]{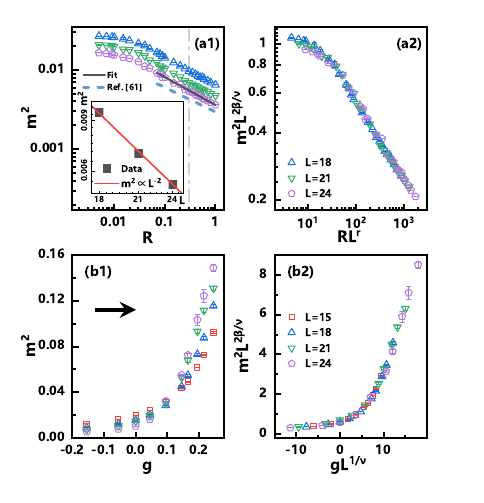}
  \vskip-3mm
  \caption{Driven dynamics from the DSM phase of model~(\ref{eq:Hamiltonian1}). \textbf{(a)} Log-log plots of $m^2$ versus $R$ for different $L$ driven to $V_c=1.355$ before (a1) and after (a2) rescaling. Inset in (a1) shows $m^2\propto L^{-2}$ for $R=0.3$ (dash-dotted line). For large $R$, power fitting for $L=24$ (black solid line) shows $m^2\propto R^{-0.353(5)}$ with the exponent close to $(2\beta-d \nu)/\nu r=-0.31(4)$ (dash line) from Ref.~\cite{Hesselmann2016prb}. \textbf{(b)} Curves of $m^2$ versus $g$ for fixed $RL^r=63.34$ and different $L$ before (b1) and after (b2) rescaling. The arrow indicates the driving direction.
  }
  \label{figure4}
\end{figure}

Next, we turn to explore the driven dynamics starting from the Mott insulator initial state and $U$ is changed as $U=U_0-R\tau$ with $U_0=11.85$. This Mott insulator state has the AFM order with transverse gapless modes. For large $R$, Fig.~\ref{figure3} (a1) shows that $m^2\propto R^{0.73(2)}$ with the exponent close to $2\beta/\nu r=0.75(2)$~\cite{Sorella2016prx} and is nearly independent of $L$. Combining this scaling relation with the usual finite-size scaling $m^2\propto L^{-2\beta/\nu}$ which is restored for small $R$, the scaling form should obey
\begin{equation}
  m^2(R,L,g)=R^{2\beta/\nu r} \mathcal{G}(RL^r,gL^{1/\nu}),
  \label{eq:scaling2}
\end{equation}
where $\mathcal{G}$ is the scaling function and $g$ is also included therein. Eq.~(\ref{eq:scaling2}) is also accordant with the conventional FTS with ordered initial state~\cite{Gong2010njp,Feng2016prb,huangyy2014prb}

We rescale curves of $m^2$ versus $R$ for various $L$ at $g=0$ according to the scaling function $m^2(R,L)=L^{-2\beta/\nu} \mathcal{G}_1(RL^{r})$ and find that the rescaled curves collapse well, which confirms Eq.~(\ref{eq:scaling2}) at $g=0$. Note that in Fig.~\ref{figure3} (a2) slight deviation appears in the large $R$ region, which may stem from the influence of high-energy modes caused by fast driving. Furthermore, Fig.~\ref{figure3} (b1) shows the curves of $m^2$ versus $g$ for an arbitrary fixed $RL^r$. The rescaled results $(gL^{1/\nu},m^2L^{2\beta/\nu})$ for various $L$ collapse into a single smooth curve, as displayed in Fig.~\ref{figure3} (b2), confirming that the driven process from AFM initial state is described by Eq.~(\ref{eq:scaling2}).

The appearance of $m^2\propto R^{2\beta/\nu r}$ reflects the fact that when $\xi_R<L$ the initial ordered magnetization domain is maintained. In this case, $\mathcal{G}(RL^r,0)$ in Eq.~(\ref{eq:scaling2}) tends to a constant and $\mathcal{G}_1(RL^{r})\sim (RL^r)^{2\beta/\nu r}$. In contrast, for small $R$ with $\xi_R>L$, the usual finite-size scaling $m^2\propto L^{-2\beta/\nu}$ is recovered, indicating that $\mathcal{G}(RL^r,0)\sim (RL^r)^{-2\beta/\nu r}$ and $\mathcal{G}_1(RL^{r})$ tends to a constant.

\begin{figure}[tbp]
\centering
  \includegraphics[width=\linewidth,clip]{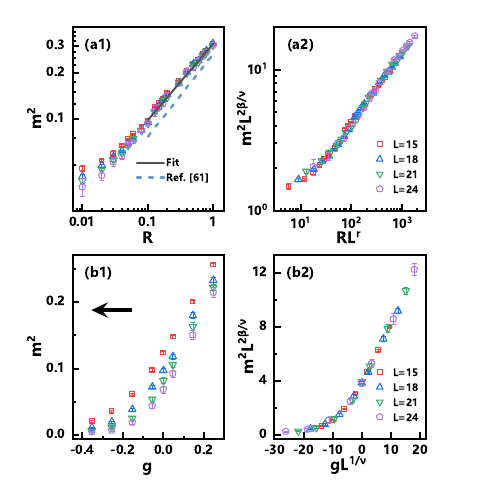}
  \vskip-3mm
  \caption{Driven dynamics from the CDW phase of model~(\ref{eq:Hamiltonian1}). \textbf{(a)} Log-log plots of $m^2$ versus $R$ for different $L$ driven to $V_c=1.355$ before (a1) and after (a2) rescaling. For large $R$, power fitting (black solid line) shows $m^2\propto R^{0.50(1)}$ with the exponent close to $2\beta/\nu r=0.54(4)$ from Ref.~\cite{Hesselmann2016prb} (dash line). \textbf{(b)} Curves of $m^2$ versus $g$ for fixed $RL^r=87.97$ and different $L$ before (b1) and after (b2) rescaling. The arrow indicates the driving direction.
  }
  \label{figure5}
\end{figure}

\noindent {\bf Dynamics in chiral Ising criticality}

To further verify the FTS in Dirac systems, we also explore the driven dynamics of Dirac QCP belonging to chiral Ising universality class, which is realized in the interacting spinless fermion model on the half-filled honeycomb lattice with the Hamiltonian~\cite{Wang2014NJP,Li2015NJP,Hesselmann2016prb}:
\begin{equation}
 H=-t\sum_{\langle ij\rangle}c_{i}^\dagger c_{j}+V\sum_{\langle ij\rangle} \left({n_{i}-\frac{1}{2}}\right) \left({n_{j}-\frac{1}{2}}\right) \label{eq:Hamiltonian1},
\end{equation}
where $V$ measures the nearest-neighbor interaction. The model is amendable to sign-problem-free QMC simulation \cite{Li2015PRB,Li2016PRL,Xiang2016PRL} (See Supplementary Materials, Sec. II). It was shown that the ground state undergoes a continuous quantum phase transition at $V_c \approx 1.355$ from the DSM phase to the charge-density-wave (CDW) insulating phase, characterized by the order parameter $m^2\equiv \sum_{i,j}\eta_i \eta_j \langle{(n_i-1/2) (n_j-1/2)}\rangle/L^{2d}$ with $\eta_i=\pm 1$ for $i$ $\in$ $A(B)$ sublattice \cite{Wang2014NJP,Li2015NJP,Hesselmann2016prb}. At $V=V_c$, both fermion and boson degrees of freedom are gapless, similar to model~(\ref{eq:Hamiltonian}). However, a difference is that in the CDW phase, the bosonic fluctuation is fully gapped owing to the discrete symmetry breaking.

For the driven dynamics under changing $V$ as $V=V_0+R\tau$ from the DSM initial state with $V_0=0$, Fig.~\ref{figure4} (a1) shows that at $g=0$, for large $R$, $m^2\propto L^{-2} R^{-0.353(3)}$ with the exponent on $R$ close to ${(2\beta-d\nu)/\nu r}=-0.31(4)$ in which $\beta=0.47(4)$, $\nu=0.74(4)$~\cite{Hesselmann2016prb}, and $z=1$~\cite{Herbut2009prb}; whereas $m^2\propto L^{-2\beta/\nu}$ for small $R$, similar to the results in model~(\ref{eq:Hamiltonian}). In addition, we find the rescaled curves of $m^2$ versus $R$ and $g$ collapse well, as shown in Fig.~\ref{figure4} (a2) and (b2), confirming that Eq.~(\ref{eq:scaling1}) gives a universal description on the driven critical dynamics from the DSM initial state.  

For the driven dynamics under changing $V$ as $V=V_0-R\tau$ from the CDW initial state with $V_0=2.5$, Fig.~\ref{figure5} (a1) shows that at $g=0$, for large $R$, $m^2\propto R^{0.50(1)}$ with the exponent close to $2\beta/\nu r = 0.54(4)$~\cite{Hesselmann2016prb}, whereas $m^2\propto L^{-2\beta/\nu}$ for small $R$, similar to the case of chiral Heisenberg universality class. Moreover, Eq.~(\ref{eq:scaling2}) is verified by the data collapse of the curves $(RL^{r},m^2L^{-2\beta/\nu})$ at fixed $g=0$ and $(gL^{1/\nu},m^2L^{-2\beta/\nu})$ for an arbitrary $RL^{r}$, as shown in Fig.~\ref{figure5} (a2) and (b2). Consequently, Eq.~(\ref{eq:scaling2}) provides a universal description on the driven dynamics from ordered initial states, regardless of whether or not initial gapless bosonic modes exist.

\noindent {\bf General scaling theory}

The above numerical results remarkably demonstrate that the FTS forms are applicable despite the existence of gapless initial states which can violate the adiabatic-impulse scenario of the original KZM. Thus, it is imperative to develop a generalized scaling scenario.

In driven dynamics starting from a ground state far from the critical point, the driving rate $R$ uniquely quantifies the extent of departure from the equilibrium state. Within the original KZM, the existence of an initial adiabatic stage stabilized by a finite gap can remarkably suppress excitations triggered by external driving~\cite{Dziarmaga2010review,Polkovnikov2011rmp,Polkovnikov2008natphy}. This leads to the fact that the excitations are predominantly generated near the QCP. Consequently, the dimension of the driving rate $R$, namely $r$, is exclusively determined by the critical exponents of the QCP, forming the basis for the original KZM and FTS.

In contrast, at first sight, for the driven dynamics evolving along a gapless initial stage and then crossing the QCP, the excitations can also be copiously produced in the initial gapless phase and subsequently brought into the critical region to influence the nonequilibrium properties near the QCP. Accordingly, we can formulate a scale transformation under which a macroscopic quantity $Q$ should transform as
\begin{eqnarray}
&&Q(R,g,L,Q_0;R')\nonumber\\
&&=b^{-\kappa}Q(Rb^r,gb^{1/\nu},Lb^{-1},Q_0b^{\kappa_0};R'b^{r'}), \label{eq:opR1}
\end{eqnarray}
in which $b$ ($b>1$) is the rescaling factor, $\kappa$ is the scaling dimension of $Q$, and $R'$ along with its exponent $r'$ characterizes the contribution from the excitations generated in the gapless initial stage. Note that although the value of $R'$ is equal to that of $R$, in general, they can have different scaling dimensions. Specifically, $r'$ is dictated by the exponent of the gapless phase; whereas $r$ is determined by the exponent of the QCP. In addition, in Eq.~(\ref{eq:opR1}), $Q_0$ and $\kappa_0$ represent the initial value of $Q$ and its scaling dimension, respectively. From Eq.~(\ref{eq:opR1}), one finds that when $r'<r$, the nonequilibrium dynamics across the QCP is governed $R$ with dimension $r$ and the usual KZM and FTS can be restored.

Moreover, in practice, the condition can be further refined. Since the gapless phase can be viewed as continuous set of critical points belonging to the same universality class, its stability requires that the tuning parameter $\Lambda$ [for example, $U$ or $V$ in the DSM phase in models~(\ref{eq:Hamiltonian}) and (\ref{eq:Hamiltonian1}), respectively] must be either irrelevant or at most marginally irrelevant. More precisely, under a scale transformation with a coarse graining factor $b$ ($b>1$), $\Lambda$ changes as $\Lambda\rightarrow \Lambda b^{\lambda}$ with $\lambda\leq 0$. Applying this scaling to a change in the tuning parameter, $\Delta \Lambda=R'\Delta t$, we obtain $\Delta \Lambda b^{\lambda}=R'b^{r'} \Delta t b^{-z'}$. This leads to the relation $r' = \lambda + z'$, where $z'$ is the dynamic exponent of the gapless phase (which may differ from $z$ characterizing QCP). Hence, $r' \leq z'$ because $\lambda \leq 0$, with equality holding for the marginally irrelevant case.

Consequently, we arrive at the precondition under which the usual KZM and FTS can be recovered for the gapless initial state,
\begin{equation}
z'<r,~~ r=z+\frac{1}{\nu}.
\label{eq:opR2}
\end{equation}
Should Eq.~(\ref{eq:opR2}) be met, the preponderant nonequilibrium universal behaviors will originate from the critical region of QCP because $r'<r$, and $R$ with exponent $r$ will govern the dynamic scaling behaviors, while the variable $R'$ can be neglected in Eq.~(\ref{eq:opR1}). To illustrate this, we consider the scaling of $m^2$. By setting the coarse-graining factor to $b=R^{-1/r}$ in Eq.~(\ref{eq:opR1}), we can transform Eq.~(\ref{eq:opR1}) into:
\begin{eqnarray}
m^2(R,g,L,m^2_0)=m^2_0R^{2\beta/\nu r-x_0/r}\mathcal{K}(gR^{-1/\nu r},LR^{1/r}), \label{eq:opR5}
\end{eqnarray}
The scaling function $\mathcal{K}(gR^{-1/\nu r},LR^{1/r})$ can be equivalently rewritten as $\mathcal{K}_1(gL^r,LR^{1/r})$. $m_0^2$ approaches a saturation value, and $x_0=0$. In this case, Eq.~(\ref{eq:opR5}) reduces to Eq.~(\ref{eq:scaling2}), thus recovering the expected scaling. In contrast, for the DSM initial state, since $m_0^2\propto L^{-d}$, $x_0=d$ and Eq.~(\ref{eq:opR5}) recovers Eq.~(\ref{eq:scaling1}).

Now we apply the general discussion elaborated above to the models we are currently dealing with. 

\begin{enumerate}[1)]
\item For driven dynamics from the DSM phase, as illustrated in Fig.~\ref{figure1}, the DSM phase is characterized by low-energy physics dominated by the fermion fluctuations exhibiting linear dispersion, $E_k=v_f |k|^{z'}$, where $z'=1$ and $v_f$ represents the fermion velocity. Given that in the DSM phase $z'=1$, Eq.~(\ref{eq:opR2}) is apparently fulfilled. Consequently, the nonequilibrium dynamics across the QCP should be described by the standard FTS form.

\item For driven dynamics from the AFM phase, in the AFM phase, the Dirac fermions feature a gap related to the amplitude of the AFM order parameter, and the low-energy excitations are predominantly attributed to bosonic spin waves, which correspond to the gapless Goldstone modes due to the broken spin rotation symmetry. The AFM spin wave has the linear dispersion as $E_k=v_b |k|^{z'}$, where $z'=1$ and $v_b$ denotes the spin wave velocity. Again, Eq.~(\ref{eq:opR2}) is satisfied and the standard FTS form should be recovered.
\end{enumerate}

Physically, in both DSM and AFM phases, different interaction strengths just yield different low-energy excitation velocity, but do not change the linear form of low-energy dispersion~\cite{doi:10.1126/science.aao2934}. Therefore, the interaction strength are marginally irrelevant and the dimension of $R'$ reduces to $z'=1$. In contrast, in the critical region of the QCP, with the onset of ordering of the bosonic order parameter fields, both the bosonic and fermionic degrees of freedom evolve into the low-energy excitations. Apart from the remarkable proliferation of low-energy modes, the mutual interactions among them trigger an intrinsic change in the scaling dimension of both Dirac fermion and order parameter, rather than merely adjusting $v_f$ and $v_b$. In this case, $U$ in model (\ref{eq:Hamiltonian}) [or $V$ in model (\ref{eq:Hamiltonian1})] becomes relevant and $(U-U_c)$ in model (\ref{eq:Hamiltonian}) [or $(V-V_c)$ in model (\ref{eq:Hamiltonian1})] has the dimension of $1/\nu$. Consequently, $R$ with dimension $r$ is more relevant than $R'$.

\noindent {\bf DISCUSSION}

In summary, we investigate the driven dynamics of QCP in two representative interacting Dirac-fermion systems, belonging to chiral Heisenberg and chiral Ising universality classes, respectively, through sign-problem-free QMC simulation.  Driving the system from both DSM and Mott insulator phases as the initial states, we discover varieties of interesting nonequilibrium scaling behaviors. Furthermore, we confirm that these scaling behaviors can be unified by the full scaling form of the FTS theory.

From the theoretical perspectives, through this work we not only successfully generalize the KZM and FTS to critical systems with joint fluctuations of gapless fermions and bosons, but also propose a general criterion for the validity of the KZM and FTS with a gapless initial state. The generalized criterion, Eq.~(\ref{eq:opR2}), can extend well beyond the scope of Dirac QCP. For instance, it was found that the density of excitations $n_{ex}$ in the driven dynamics of the one dimensional spin chain with dispersion $E_k\propto\sqrt{c^2k^2+k^4}$ shows the scaling relation $n_{ex}\propto R^{1/3}$ under changing $c$ linearly to cross its QCP at $c=0$~\cite{PhysRevB.78.144301}. In this case, the low-energy mode of the gapless phase has the linear dispersion with $z'=1$ and $c$ is marginally irrelevant with zero scaling dimension. In contrast, near the QCP, the quadratic dispersion begins to dominate such that $z=2$, and $c$ becomes relevant with the scaling dimension of $1/\nu=1$, which can be obtained by comparing the dimensions of $c$ and $k$. Apparently, the criterion of Eq.~(\ref{eq:opR2}) is fulfilled and $n_{ex}$ satisfy the KZM of the QCP, i.e., $n\propto R^{d/r}$ with $d=1$ and $r=3$. In addition, Eq.~(\ref{eq:opR2}) can also apply to the dynamic scaling in other QCPs~\cite{PhysRevB.92.064419,Deng2008epl},  as discussed further in Supplementary Materials, Sec. III.

From the application perspectives, we demonstrate that the nonequilibrium scaling form is capable of determining the critical exponents in Dirac QCP, providing an effective method to deciphering quantum critical properties in terms of driven dynamics. A large discrepancy in the critical exponents of the chiral Heisenberg universality class still exists despite extensive studies~\cite{PhysRevB.91.165108,Sorella2016prx}. By directly fitting the curves of $m^2$ versus $R$ in large $R$ region with the power function for both DSM and AFM initial states, as shown in Fig.~\ref{figure2}(a1) and Fig.~\ref{figure3}(a1), respectively, we obtain $(2\beta/\nu r-d/r)=-0.26(1)$ and $2\beta/\nu r=0.73(2)$. By setting $z=1$, we obtain critical exponents $\nu=0.98(4)$ and $\beta=0.72(4)$, which are consistent with $\nu=1.02(1)$ and $\beta=0.76(2)$ from Ref.~\cite{Sorella2016prx} within error bars. The detailed derivation and other ways to estimate critical exponents via FTS are shown in the Supplementary Materials, Secs. IV and V. It is important to note that equilibrium methods often require larger system sizes and careful finite-size scaling corrections to achieve comparable accuracy~\cite{Sorella2016prx}. In contrast, in the framework of driven dynamics, the driven rate $R$ constitutes a new tuning parameter. At large $R$, the impact of finite-size corrections is diminished, allowing for reliable critical exponent determination with relatively small system sizes. Moreover, the availability of diverse driving protocols with distinct FTS forms provides a compelling means to confirm the robustness and accuracy of these critical exponent determinations  (See Supplementary Materials, Sec. V.).

From experimental perspectives, driven dynamics is observed in cold-atom systems~\cite{Clark2016science,Keesling2019}. Due to recent developments in cold-atom-based quantum simulators of fermions~\cite{Esslinger2008Nature,Greiner2017Nature,Venu2023Nature}, it is promising to detect driven dynamics in Dirac QCP and verify the generalized KZM and FTS as discussed in our study in these platforms. In real experiments, thermal fluctuations will inevitably enter. Thus, the scaling form should include $TR^{-z/r}$ as an additional variable in the full scaling form~\cite{Yin2014prb,PhysRevB.94.064302}. Nonetheless, as $T^{-1/z}$ plays similar roles as the system size in QCP, the scaling behaviors for large $R$ are expected to be almost independent of $T$, in analogy to the finite-size cases. Physically, for large $R$, excitations induced by driving dominate over those induced by thermal fluctuations and therefore contribute the main dynamic scaling behaviors. Accordingly, it is foreseeable that the methodological approach developed here can be used to probe the critical properties in real experiments.

\noindent {\bf METHODS}

{\bf Quantum Monte Carlo simulation of driven dynamics in imaginary-time direction}---We have explored the driven dynamics of the Hubbard model \eqref{eq:Hamiltonian} and the $t$-$V$ model \eqref{eq:Hamiltonian1}. Here, we elucidate the implementation of QMC to simulate the imaginary-time driven dynamics.

For model \eqref{eq:Hamiltonian} and \eqref{eq:Hamiltonian1}, we linearly vary the interaction strength $U~(V)$ with imaginary time variable $\tau$ at the rate $R$ as 
    \begin{equation}
        U(\tau) = U_0 \pm R \tau
        \label{eq:varyU},
    \end{equation}
    \begin{equation}
        V(\tau) = V_0 \pm R \tau
        \label{eq:varyV},
    \end{equation}
    where the $+(-)$ represents driving from disordered (ordered) initial state at $U_0~(V_0)$. 
    In our numerical simulation, for disordered initial state, we set $U_0=0,~V_0=0$, while for ordered initial state we set $U_0=11.85,~V_0=2.5$ far from the critical point.
    
    The wave function $|\psi(\tau)\rangle$ obeys the imaginary-time Schr\"{o}dinger equation~\cite{Polkovnikov2011prb}
    \begin{equation}
     -\frac{\partial}{\partial \tau} |\psi(\tau)\rangle=H(\tau)|\psi(\tau)\rangle
    \label{eq:dynamiceq}.
    \end{equation}
    The formal solution of Eq.~(\ref{eq:dynamiceq}) is given by $|\psi(\tau)\rangle=U(\tau,0)|\psi(0)\rangle$ in which time evolution operator 
    \begin{equation}
        U(\tau,0) =\mathrm{T}~ {\rm exp}\left[-\int_{0}^{\tau}d\tau'H(\tau')\right],
    \end{equation}
    with $\mathrm{T}$ being the time-ordering operator in imaginary-time direction. For the left vector $\bra{\psi (\tau)}=\bra{\psi (0)}U^\dagger (\tau,0)$ with
    $
        U^\dagger(\tau,0) =\overline{\mathrm{T}}~ {\rm exp}\left[-\int_{0}^{\tau}d\tau'H(\tau')\right],
    $
    the Hermite conjugate simply changes the time-ordering operator $\mathrm{T}$ to an anti-time-ordering operator $\overline{\mathrm{T}}$.
    Since the model~(\ref{eq:Hamiltonian}) and~(\ref{eq:Hamiltonian1}) is sign-problem-free and imaginary-time evolution does not induce additional sign problem, the imaginary-time dynamics of the models can be simulated by the determinant quantum Monte Carlo (DQMC) method.

    To facilitate DQMC simulations of models~\eqref{eq:Hamiltonian} and~\eqref{eq:Hamiltonian1}, we begin by expressing the Hamiltonian in the form
    \begin{equation}
        H(\tau) = H_t + H_I(\tau),
    \end{equation}
    where $H_t$ represents the kinetic energy and $H_I(\tau)$ the interaction term, in which the interaction strength varies with $\tau$. The initial state $|\psi(0)\rangle$ is the ground state of $H(0)$. In the DQMC simulation~\cite{AssaadReview}, the ground state of a given Hamiltonian is accessed by performing imaginary-time evolution on a trial wave function:
    \begin{equation}
    |\psi(0)\rangle=\lim_{\tau_0 \rightarrow \infty} e^{-\tau_0  H(0)} | \psi_{T} \rangle, 
    \label{eq:initial}
    \end{equation}
    where $| \psi_{T} \rangle$ is a Slater-determinant wave function generated as the ground state of the $H_t$. Consequently, for disordered DSM initial state $ |\psi(0)\rangle=|\psi_{T} \rangle$, we set projection time $\tau_0 = 0$. For the AFM or CDW ordered initial state, a sufficiently long $\tau_0$ is necessary to project the $|\psi_{T}$ onto the ground state of $H(0)$. In our simulations, we implement $\tau_0 = 120$ for the model \eqref{eq:Hamiltonian}, and $\tau_0 = 20$ for the model \eqref{eq:Hamiltonian1}. We have numerically verified the convergence of our results with respect to increasing $\tau_0$.

    Following the standard DQMC methodology~\cite{AssaadReview}, we employ the Trotter decomposition to discretize imaginary time. Subsequently, we apply the Hubbard-Stratonovich (HS) transformation to decouple the fermion-fermion interaction, transforming it into a fermion bilinear form coupled to auxiliary fields. First, we perform Trotter decomposition on the process of imaginary-time evolution to generate initial state in \eqref{eq:initial}: 
    \begin{equation}
    |\psi(0)\rangle= \lim_{\Delta\tau \rightarrow 0}\prod_{n=1}^{N_{\tau_0}} e^{-\Delta\tau  H_t} e^{-\Delta\tau  H_I(0)} | \psi_{T} \rangle, 
    \label{eq:initial2}
    \end{equation}
    where $\Delta\tau$ is the imaginary-time Trotter time defined by $\Delta\tau = \tau_0/N_{\tau_0}$. 
    
    Similarly, the Trotter decomposition in driven dynamics is expressed as 
    \begin{equation}
        \mathrm{T}~ {\rm exp}\left[-\int_{0}^{\tau}d\tau'H(\tau')\right] = \lim_{\Delta\tau \to 0} \prod^{N_\tau}_{n=1} \left[\mathrm{e}^{-\Delta\tau H_t} \mathrm{e}^{-\Delta\tau H_I(\tau_n)} \right],
        \label{eq:dynamics2}
    \end{equation}
    where the Trotter time is defined as $\Delta\tau=\tau/N_\tau$, with with $N_\tau$ the number of time slices and $\tau_n=n\Delta\tau$.  Combining these two parts, the wave function $|\psi(\tau)\rangle$ is expressed as:
    \begin{equation}
    |\psi(\tau)\rangle = U(\tau,0)P_G | \psi_{T} \rangle,
        \label{eq:dynamics3}
    \end{equation}
where $U(\tau,0)= \lim_{\Delta\tau \to 0} \prod^{N_\tau}_{n=1} \left[\mathrm{e}^{-\Delta\tau H_t} \mathrm{e}^{-\Delta\tau H_I(\tau_n)}\right] $ and $P_G = \lim_{\Delta\tau \to 0} \prod_{n=1}^{N_{\tau_0}} \left[e^{-\Delta\tau  H_t} e^{-\Delta\tau  H_I(0)}\right] $. In our practical implementation, we choose $\Delta\tau=0.05$ in both \eqref{eq:initial2} and \eqref{eq:dynamics2}. We have numerically confirmed that $\Delta\tau = 0.05$ is sufficiently small to guarantee that the time-step error introduced by the Trotter decomposition is negligible, with the details shown in the Supplementary Materials, Sec. VI.

Next, we apply the HS transformation to decouple the interacting part $H_I(\tau)$ in models \eqref{eq:Hamiltonian} and \eqref{eq:Hamiltonian1}. For the Hubbard interaction in \eqref{eq:Hamiltonian}:
    \begin{equation}
    \begin{aligned}
        & \mathrm{e}^{-\frac{\Delta\tau U(\tau_n)}{2}\left(n_{i\uparrow}+n_{i\downarrow}-1\right)^2}
        \\
        &=
        \sum_{l_{i,\tau_n}=\pm 1,\pm 2}
        \gamma(l_{i,\tau_n})
        \mathrm{e}^{\mathrm{i} \sqrt{\frac{\Delta\tau U(\tau_n)}{2}}\eta(l_{i,\tau_n})\left(n_{i\uparrow}+n_{i\downarrow}-1\right)},
        \label{seq:HS}
    \end{aligned}
    \end{equation}
    and for the density interaction in \eqref{eq:Hamiltonian1}:
    \begin{equation}
    \begin{aligned}
        &\mathrm{e}^{-\frac{\Delta\tau V(\tau_n)}{2}\left(c^{\dagger}_{i}c_{j}+c^{\dagger}_{j}c_{i}\right)^2}
        \\
        &=
        \sum_{l_{i,\tau_n}=\pm 1,\pm 2}
        \gamma (l_{i,\tau_n})
        \mathrm{e}^{\mathrm{i} \sqrt{\frac{\Delta\tau V(\tau_n)}{2}}\eta(l_{i,\tau_n})\left(c^{\dagger}_{i}c_{j}+c^{\dagger}_{j}c_{i}\right)},
        \label{seq:isingHS}
    \end{aligned}
    \end{equation}
    where the four-component space-time auxiliary fields $\gamma$ and $\eta$ take the following values:
    \begin{equation}
         \gamma(\pm1)=1+\sqrt{6}/3, \
         \gamma(\pm2)=1-\sqrt{6}/3,
    \end{equation}
    \begin{equation}
         \eta(\pm1)=\pm\sqrt{2\left(3-\sqrt{6}\right)}, \
         \eta(\pm2)=\pm\sqrt{2\left(3+\sqrt{6}\right)}.
    \end{equation}

    Subsequent to the procedures outlined above, both the imaginary-time evolution operator $U(\tau,0)$ and the generator of the initial state $P_G$ can be expressed as products of exponentials of fermion bilinear operators.  This representation significantly simplifies calculations, enabling the straightforward determination of wavefunction overlaps under imaginary-time evolution $\left\langle \psi(\tau) \middle| \psi(\tau) \right\rangle$ and the expectation values of observables $\braket{O(\tau)} = 
        \frac{
                \bra{\psi(\tau)}O\ket{\psi(\tau)}
            }{
                \left\langle \psi(\tau) \middle| \psi(\tau) \right\rangle\
            }$ in the framework of conventional DQMC. For a comprehensive exposition of these techniques, we refer the reader to earlier works~\cite{AssaadReview}.

    For updating the auxiliary fields, we employ a local update scheme, sequentially updating the field at each site and time slice. In non-equilibrium simulations, one Markov chain consists of $N_{\mathrm{sweep}}$ iterations.  Each iteration involves updating all sites across all time slices, resulting in a total of $N_{\mathrm{sweep}}(N_{\tau}+N_{\tau_0})N_{\mathrm{site}}$ Monte Carlo updates per chain, where $N_{\mathrm{sweep}}$ typically ranges from $10^2$ to $10^3$.  Here, $N_{\mathrm{site}}$ represents the number of lattice sites.   To ensure thermalization, we perform an initial equilibration of $5$ sweeps before taking measurements.  Finally, for each data point, we typically run two independent Markov chains to improve statistical sampling.

\noindent {\bf Data availability}

The data that support the findings of this study are available from the corresponding authors (Z.X.L. and S.Y.) upon request.

\noindent {\bf Code availability}

All numerical codes in this paper are available upon request to the corresponding authors (Z.X.L. and S.Y.).

{\bf Acknowledgments}

We would like to thank A. W. Sandvik and F. Zhong for helpful discussions. Z. Zeng, Y. K. Yu, Z. X. Li and S. Yin are supported by the National Natural Science Foundation of China (Grants No. 12222515 and No. 12075324). Z. X. Li is supported by the NSFC under Grant No. 12347107. S. Yin is also supported by the Science and Technology Projects in Guangdong Province (Grant No. 2021QN02X561) and Guangzhou City (Grant No. 2025A04J5408).

{\bf Author contributions}

S.Y. and Z.X.L conceived the project and planned the study. The numerical simulations
were carried out by Z.Z., Y.Y.K. and Z.X.L. All authors contributed to the scaling analyses.

\onecolumngrid
\newpage
\widetext
\thispagestyle{empty}

\setcounter{equation}{0}
\setcounter{figure}{0}
\setcounter{table}{0}
\renewcommand{\theequation}{S\arabic{equation}}
\renewcommand{\thefigure}{S\arabic{figure}}
\renewcommand{\thetable}{S\arabic{table}}

\pdfbookmark[0]{Supplementary Materials}{SM}
\begin{center}
    \vspace{3em}
    {\Large\textbf{Supplementary Materials for}}\\
    \vspace{1em}
    {\large\textbf{Finite-time scaling beyond the Kibble-Zurek prerequisite in Dirac systems}}\\
    \vspace{0.5em}
\end{center}

\section{I. Consistency of scaling in imaginary-time and real-time driven dynamics
}

In the context of driven critical dynamics that starts from a ground state far away from the critical point, the driving rate $R$ serves as the only quantity to measure the extent to which the system departs from its equilibrium state for both real-time and imaginary-time cases. In addition, both the real time and imaginary time have the same scaling dimension $z$. Accordingly, the critical dimension of $R$ should be the same for both real-time and imaginary-time driven dynamics.

To see this explicitly, one can expand the evolving wave function on the basis of instantaneous eigenstates of the Hamiltonian and compare the coefficients of the excited states under linearly driving in real-time $t$ direction and imaginary-time $\tau$ direction. The initial state is always assumed to be the ground state of the initial parameter $\Lambda_0$ that is far from the critical point.

For the driven dynamics in real-time direction, the coefficient for the $n$-th excited state is~\cite{PhysRevB.72.161201}
\begin{equation}
  a_n(t)\simeq -\int_{\Lambda_0}^{\Lambda_f} d\Lambda' \langle n|\partial_{\Lambda'}|0\rangle \exp\left[i\int_{\Lambda_0}^{\Lambda'}d\Lambda'' \frac{\Delta_{n0}(\Lambda'')}{R}\right],
\label{realt}
\end{equation}
in which $\Delta_{n0}(g)\equiv E_n(g)-E_0(g)$ is energy difference between the $n$-th eigenstate and the ground state of the instantaneous Hamiltonian $H(g)$. In contrast, for the imaginary-time driven dynamics, the coefficient for the $n$-th excited state is~\cite{Polkovnikov2011prb,De_Grandi_2013}
\begin{equation}
  \alpha_n(t)\simeq \int_{\Lambda_0}^{\Lambda_f} d\Lambda' \langle n|\partial_{\Lambda'}|0\rangle \exp\left[-\int_{\Lambda_0}^{\Lambda'}d\Lambda'' \frac{\Delta_{n0}(\Lambda'')}{R}\right].
\label{imaginarytau}
\end{equation}

By comparing Eqs.~(\ref{realt}) and (\ref{imaginarytau}), one finds that the difference is the constant coefficient of the argument in the exponential term, which is imaginary unit for real-time dynamics, but unit for imaginary-time dynamics. However, universal scaling behaviors are captured by other common variables: including the distance to the critical point $g=\Lambda-\Lambda_c$, the transition matrix $\langle n|\partial_{\Lambda}|0\rangle$, the energy difference $\Delta_{n0}(\Lambda)$, and the driving velocity $R$. Accordingly, the scaling theory for both real-time and imaginary-time dynamics should share a similar scaling form, with the same critical exponents, differing only in the specific details encoded within the scaling function.

\section{II. Imaginary-time quantum Monte Carlo and sign problem}

The sign problem remains a major obstacle to performing efficient QMC simulations for large systems and at low temperatures.  Nevertheless, specific classes of models are known to be sign-problem-free, and time-reversal symmetry (TRS) is frequently exploited within determinant QMC to establish this property. In the honeycomb Hubbard model we consider (Eq.~(\ref{eq:Hamiltonian}) in the maintext), the absence of the sign problem is guaranteed by the presence of standard time-reversal symmetry in the complex fermion basis.  Crucially, following the Hubbard-Stratonovich transformation, the model acquires a TRS with \(\mathcal{T}^2=-1\) in spin space, which ensures the sign-problem-free condition.  For the honeycomb \(t\)-\(V\) model (Eq.~(\ref{eq:Hamiltonian1}) in the maintext), the sign-problem-free characteristic is attributed to the existence of two distinct time-reversal symmetries acting within the Majorana-fermion space.  According to the sign-free QMC principles introduced in Ref.~\cite{Li2016PRL}, this places the model within the Majorana class and guarantees the absence of the sign problem. The details of the sign-free conditions are provided in the review article~\cite{Li2019Review}.

For the real-time dynamics of the fermion model, in addition to the sign problem in the model itself, there is another source of sign problem.  The real-time evolution operator, $\exp{(-i H t)}$, introduces a``complex" factor to the wavefunction, which manifests as a sign problem in Monte Carlo simulations. Consequently, even if a model is sign-problem-free in its static formulation, simulating its real-time dynamics is generally sign-problematic.

In contrast,  if a model is sign-problem-free, the imaginary-time dynamics simulation will also remain free from the sign problem. This is because the imaginary-time evolution operator, $\exp{(-H t)}$, does not introduce any additional complex factors.  Consequently, for sign-problem-free models, imaginary-time dynamics can be reliably simulated using quantum Monte Carlo methods.

\section{III. Other evidences supporting the general criterion}
The general criterion for the KZM and FTS with gapless initial state discussed in the main text can be further corroborated by the adiabatic perturbation calculations on the density of excitations, $n_{ex}=(1/L^d)\sum_n |\alpha_n|^2$, in which $\alpha_n$ denotes the coefficient for the $n$-th excited state~\cite{PhysRevB.72.161201}. Here we focus on the imaginary-time dynamics. Same scaling relations can be obtained for the real-time case with the similar procedure.

According to the discussions in the main text, in both the DSM and AFM phases, the linearly driving force can be regarded as being applied to the coefficient of linear dispersion. In this case, adiabatic perturbation calculations show that $n_{ex}$ can be approximated as

\begin{equation}
n_{ex}\approx \int \frac{d^dk}{(2\pi)^d}\left|\int_{\Lambda_i}^{\Lambda_f} d\Lambda \langle k|\partial_{\Lambda}|0\rangle \exp\left[-\frac{k^{z'}}{R'}\int_{\Lambda_i}^{\Lambda}d\Lambda' v_{f/b}(\Lambda')\right]\right|^2, \label{eq:opR3}
\end{equation}
in which $\Lambda_i$ and $\Lambda_f$ correspond to the parameters in the same phases. Here the momentum $k$, measuring the momentum difference between the excited state and the ground state, is used to denote the energy levels. By rescaling $k$ as $k=R'^{1/z'}\kappa$ in Eq.~(\ref{eq:opR3}), one finds that 
\begin{equation}
n_{ex}\approx R'^{d/z'}\int \frac{d^d\kappa}{(2\pi)^d}\left|\int_{\Lambda_i}^{\Lambda_f} d\Lambda \langle \kappa R'^{1/z'}|\partial_{\Lambda}|0\rangle \exp\left[-\kappa^{z'}\int_{\Lambda_i}^{\Lambda}d\Lambda' v_{f/b}(\Lambda')\right]\right|^2. \label{eq:opR31}
\end{equation}
Thus one obtains $n_{ex}\propto R'^{d/z'}$ with the coefficient being an integral of $\kappa$~\cite{Polkovnikov2008natphy}.

In contrast, for the driving dynamics across the QCP, 
\begin{equation}
n_{ex}\approx \int \frac{d^dk}{(2\pi)^d}\left|\int_{\Lambda_0}^{\Lambda_f} d\Lambda' \langle k|\partial_{\Lambda'}|0\rangle \exp\left[-\int_{\Lambda_0}^{\Lambda'}d\Lambda'' \frac{\Delta_{k0}(\Lambda'')}{R}\right]\right|^2,
\label{eq:opR32}
\end{equation}
in which $\Lambda_i$ and $\Lambda_f$ belong to the different phases. Near the critical point with a small $g$, $\Delta_{k0}(g)=g^{\nu z}F_{\Delta}(kg^{-\nu})$, and the matrix element $\langle k|\partial_{g}|0\rangle=g^{-1}F_{M}(kg^{-\nu})$. By substituting these two scaling relations into Eq.~(\ref{eq:opR32}) and rescaling $g$ and $k$ as $g=\gamma R^{1/\nu r}$ and $k=\kappa R^{1/r}$, respectively, in which $r=z+1/\nu$, one obtains~\cite{PhysRevB.72.161201}
\begin{equation}
n_{ex}\approx R^{d/r}\int \frac{d^d\kappa}{(2\pi)^d}\left|\int_{\gamma_i}^{\gamma_f} d\gamma {\gamma}^{-1}F_M(\kappa \gamma^{-\nu}) \exp\left[-\int_{\gamma_i}^{\gamma}d\gamma' \gamma'^{\nu z}F_{\Delta}(\kappa \gamma'^{-\nu})\right]\right|^2. \label{eq:opR33}
\end{equation}
Accordingly, we have $n_{ex}\propto R^{d/r}$.

By comparing these two scaling relations, $n_{ex}\propto R'^{d/z'}$ and $n_{ex}\propto R^{d/r}$, we find that when $z'<r$, the excitation is dominated by the QCP for the KZM dynamics in which the driving rate is not very large, consistent with above scaling analyses. Note that in practice, our numerical results show that the applicable range of the driving rate can even approach one, indicating that the coefficients before the scaling relation of $n_{ex}\propto R'^{d/z'}$ is much smaller than that of $n_{ex}\propto R^{d/r}$. Actually, it was shown that $v_f$ exhibits only a weak dependence on the interaction term over a broad range of interaction strengths~\cite{doi:10.1126/science.aao2934}. This property narrows down the integral range of $\Lambda$ that furnishes the major contribution in Eq.~(\ref{eq:opR3}) and thereby reducing the coefficient.

Moreover, we show that the general criterion of Eq.~(\ref{eq:opR2}) can also apply in the driven dynamics of the QCP separating the gapless Tomonaga-Luttinger phase and the ferromagnetic phase in one dimension~\cite{PhysRevB.92.064419}. In the gapless Tomonaga-Luttinger liquid phase, $z'=1$. In contrast, near the critical point, $z=2$ and $1/\nu=2$. Accordingly, For driven dynamics from the Tomonaga-Luttinger phase to cross the QCP, the criterion of Eq.~(\ref{eq:opR2}) is fulfilled. Consequently, the QCP dominate the dynamic critical properties. Note that the consistent conclusion was obtained in Ref. \cite{PhysRevB.92.064419}. However, the explanation is different.

A marginal example appears in the spin-$1/2$ XY chain with an alternating transverse field~\cite{Deng2008epl}. The driven dynamics by changing the parameter $\delta$ along a gapless line in the phase diagram to cross the critical point with $\delta=0$ was considered. The gap function is $E_k=4 \sqrt{k^2 + 2 \delta (\delta - \sqrt{k^2 + \delta^2})}$. When $\delta\neq 0$, the low energy spectra of $E_k$ have the quadratic form and $z'=2$. (Note that in Ref.~\cite{Deng2008epl}, a critical exponent was identified as the correlation length exponent of the gapless phase. However, this exponent does not relate to the variable $\delta$. So, this exponent should play no role for changing $\delta$.) Near the critical point, the dimension of $\delta$ is $1/\nu=1$. At $\delta=0$, $E_k$ reduces to $E_k=4|k|$ and $z=1$. In this case, $z'=z+1/\nu$. In Ref.~\cite{Deng2008epl}, numerical simulation shows that $n_{ex}\propto R^{1/2}$. This scaling relation was explained as the dynamic scaling dominated by the QCP. However, according to the scenario developed in present paper, this scaling behavior should correspond to the marginal case, in which the defects generated in the gapless phase and in the critical region have the same scaling relation.

\section{IV. RESULTS OF DIMENSIONLESS QUANTITY}
In the main text, we only present dynamic results of the square of the order parameter $m^2$. Here, we supplement the results of a dimensionless quantity correlation ratio defined as~\cite{PhysRevLett.115.157202}:
\begin{equation}
	R_{S}\equiv{1-S(\Delta \bm q)}/{S(\bm 0)},
	\label{eq:RS}
\end{equation}
where $S({\bm q})$ is the structure factor and $\Delta\bm{q}$ is the minimal lattice momentum.

\begin{figure}[htbp]
	\centering
	\includegraphics[width=\linewidth]{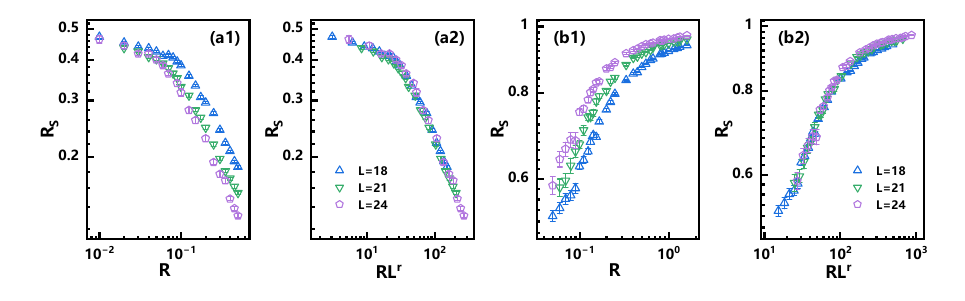}
	\vskip-3mm
	\caption{Log-log plots of $R_S$ versus $R$ for different $L$ driven to $U_c=3.85$, for the honeycomb Hubbard model. \textbf{(a)} displays the results driven from the DSM phase before (a1) and after (a2) rescaling. \textbf{(b)} displays the results driven from the AFM phase before (b1) and after (b2) rescaling.}
	\label{figure_chiralHeisenberg_RS}
\end{figure}

\begin{figure}[htbp]
	\centering
	\includegraphics[width=\linewidth]{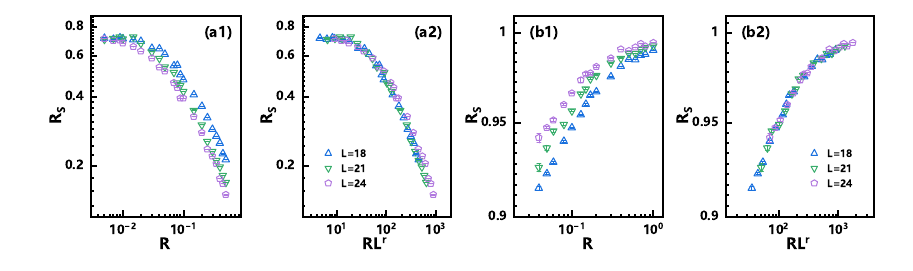}
	\vskip-3mm
	\caption{Log-log plots of $R_S$ versus $R$ for different $L$ driven to $V_c=1.355$, for the $t-V$ model. \textbf{(a)} displays the results driven from the DSM phase before (a1) and after (a2) rescaling. \textbf{(b)} displays the results driven from the CDW phase before (b1) and after (b2) rescaling.}
	\label{figure_chiralIsing_RS}
\end{figure}

For the honeycomb Hubbard model exhibiting phase transition belonging to the chiral Heisenberg universality class, $S({\bm q})$ is the antiferromagnetic structure factor:
\begin{equation}
	S({\bm{q}})_{\rm AFM}=  \frac{1}{L^{2d}} \sum_{i,j} \mathrm{e}^{\mathrm{i} \bm{q}\cdot ({\bm{r}_i-\bm{r}_j})} \langle{S_i^z S_j^z}\rangle,
	\label{eq:S_AFM}
\end{equation}
where the staggered magnetization operator $S_i^{z}\equiv\vec{c}^\dagger_{i,A}\sigma^z\vec{c}_{i,A}-\vec{c}^\dagger_{i,B}\sigma^z\vec{c}_{i,B}$ with $i$ characterizes cells and ${\vec{c}^\dagger\equiv({c_\uparrow^\dagger,c_\downarrow^\dagger}})$.

For the honeycomb $t$-$V$ model exhibiting phase transition belonging to the chiral Ising universality class,  $S({\bm q})$ is the charge-density-wave structure factor:
\begin{equation}
	S({\bm{q}})_{\rm CDW}= \frac{1}{L^{2d}} \sum_{i,j} \mathrm{e}^{\mathrm{i} \bm{q}\cdot ({\bm{r}_i-\bm{r}_j})} \langle{n_i n_j}\rangle,
	\label{eq:S_CDW}
\end{equation}
where $n_i\equiv n_{i,A} - n_{i,B}$ with $n_{i,A}$ being the electron number operator defined as $n_{i,A}\equiv c^\dagger_{i,A}c_{i,A}$.

According to the scaling analysis in the main text, as a dimensionless variable, $R_{S}$ in the driven process obeys the following dynamic scaling form:
\begin{equation}
	R_{S}({g,R,L})=f(RL^r,gL^{1/\nu}),
	\label{eq:RS_scale}
\end{equation}
which indicates that \eqref{eq:RS_scale} at $g=0$ can be converted to $R_S = f(RL^r)$.

As shown in Fig.~\ref{figure_chiralHeisenberg_RS}(a2)(b2) and Fig.~\ref{figure_chiralIsing_RS}(a2)(b2), we find that the rescaled curves  collapse well with the known critical exponents set as input~\cite{Sorella2016prx,Hesselmann2016prb}. These results not only confirm \eqref{eq:RS_scale} at $g=0$, but also support the conclusions we draw in the main text. Moreover, the scaling behaviors of these dimensionless quantities provide alternative methods to estimate the critical exponents, as discussed in the following section.

\section{V. Estimation OF CRITICAL EXPONENTS}

The finite-time scaling (FTS) provides powerful approach to revealing critical properties. Here, we show various ways to estimate critical exponents using the scaling forms of FTS.

\textbf{Method 1:} For the honeycomb Hubbard model, given the critical point $U_c=3.85$~\cite{Sorella2016prx}, the data fitting of $m^2$ versus $R$ corresponding to $m^2\propto L^{-d} R^{(2\beta-d\nu)/\nu r}$ and $m^2\propto R^{2\beta/\nu r}$, the leading terms in the FTS scaling form, gives $(2\beta-d\nu)/\nu r = -0.26(1) = k_{\mathrm{DSM}}$ and $2\beta/\nu r = 0.73(2) = k_{\mathrm{AFM}}$, where $r=z+1/\nu$. In our models, $d=2$ and $z=1$. Subtracting the two leading terms yields an expression solely in terms of the critical exponent $\nu$, $\nu = 2 / (2+k_{\mathrm{DSM}}-k_{\mathrm{AFM}}) - 1$. Accordingly, we obtain $\nu=0.98(4)$. Combining the values of $\nu$ and $k_{\mathrm{DSM}}$ gives rise to $\beta=0.72(4)$.

\textbf{Method 2:} Using data collapse of the results to determine the critical properties is another approach. For instance, to obtain $\nu$, we consider the correlation ratio $R_S$ at $U_c=3.85$. Since at the critical point, Eq.~\eqref{eq:RS_scale} is converted to $R_S = f(RL^r)$. Given $z=1$, the data collapse gives $\nu=1.04(5)$. Then based on the data collapse of order parameter according to the scaling form $m^2=L^{-2\beta/\nu} \mathcal{F}(RL^{r},0)$, we obtain $\beta = 0.80(4)$.

\textbf{Method 3:} The combination of the fitting of leading terms and the data collapse provides various approaches for evaluating critical exponents. For instance, combining $\nu = 1.04(5)$ obtained in Method 2 and the slope $k_{\mathrm{DSM}}=-0.26(1)$, gives $\beta = 0.78(5)$. 

As shown in Table~\ref{tab:heisenberg}, the results obtained from the scaling forms of FTS are not only consistent with the critical exponents $\nu=1.02(1)$ and $\beta=0.76(2)$ from Ref.~\cite{Sorella2016prx}, but also mutually consistent, within the error bars.

\begin{table}[htbp]
	\centering
	\caption{Critical exponents of the chiral Heisenberg universality class obtained from diverse methods. The first three rows of data present the results calculated using the three methods listed above. The remaining rows display related research findings for comparison purposes. In the table, 'honeycomb' refers to the Hubbard model on a honeycomb lattice.}
	\vspace{1em}
	\begin{tabularx}{\textwidth}{l *{3}{>{\centering\arraybackslash}X}}
		\toprule
        \textbf{Model} &
		\textbf{Method} & \textbf{$\bm{\nu}$} & \textbf{$\bm{\beta}$} \\
		\midrule
		honeycomb & QMC-FTS-Method 1 (present) & 0.98(4) & 0.72(4)  \\
        honeycomb & QMC-FTS-Method 2 (present) & 1.04(5) & 0.80(4)  \\
		honeycomb & QMC-FTS-Method 3 (present)& 1.04(5) & 0.78(5)  \\
		honeycomb & QMC~\cite{Sorella2016prx} & 1.02(1) & 0.76(2) \\
        honeycomb & QMC~\cite{PhysRevB.91.165108} & 0.84(4) & 0.71(8) \\
        Gross-Neveu & $4-\epsilon$, 1st order~\cite{ROSENSTEIN1993381} & 0.851 & 0.804 \\
        Gross-Neveu & $4-\epsilon$, 2nd order~\cite{ROSENSTEIN1993381} & 1.01 & 0.995 \\
        Gross-Neveu & FRG~\cite{PhysRevB.89.205403} & 1.31 & 1.32\\
		\bottomrule
	\end{tabularx}
	\label{tab:heisenberg}
\end{table}

\begin{table}[htbp]
	\centering
	\caption{Critical exponents of the chiral Ising universality class obtained from diverse methods, following the same convention as in Table~\ref{tab:heisenberg}.}
	\vspace{1em}
	\begin{tabularx}{\textwidth}{l *{3}{>{\centering\arraybackslash}X}}
		\toprule
		\textbf{Model} & 
        \textbf{Method} &
        \textbf{$\bm{\nu}$} & \textbf{$\bm{\beta}$} \\
		\midrule
		honeycomb & QMC-FTS-Method 1 (present) & 0.74(2) & 0.43(1)  \\
        honeycomb & QMC-FTS-Method 2 (present) & 0.74(2) & 0.46(1)  \\
		honeycomb & QMC-FTS-Method 3 (present) & 0.74(2) & 0.43(1)  \\
		honeycomb & QMC~\cite{Hesselmann2016prb} & 0.74(4) & 0.47(4) \\
        honeycomb & QMC~\cite{Li2015NJP} & 0.77(3) & 0.60(3) \\
        honeycomb & QMC~\cite{Wang2014NJP} & 0.80(3) & 0.52(2) \\
        Gross-Neveu & $4-\epsilon$, 1st order~\cite{ROSENSTEIN1993381} & 0.709 & 0.559 \\
        Gross-Neveu & $4-\epsilon$, 2nd order~\cite{ROSENSTEIN1993381} & 0.797 & 0.610 \\
        Gross-Neveu & FRG~\cite{rosa2001} & 0.927 & 0.707\\
		\bottomrule
	\end{tabularx}
	\label{tab:ising}
\end{table}

For the honeycomb lattice $t-V$ model, a set of methods similar to those for the honeycomb Hubbard model is used to estimate the exponent values, which is shown in Table~\ref{tab:ising}. The exponents obtained through various ways using FTS align not only with those from Ref.~\cite{Hesselmann2016prb} but also with each other, within the margin of error.

Using the leading terms in the FTS scaling forms to estimate critical exponents is a feature of our method, which can effectively eliminate the influence of finite-size effects for large driving rate. In addition, there are other ways to perform data collapse, such as using the off-critical-point effects in the FTS forms.

In summary, the above derivation and results demonstrate that our method is reliable and promising for investigating quantum critical phenomena in fermionic systems.

\section{VI. THE ANALYSIS OF TIME STEP ERROR}

In DQMC simulations, the time step $\Delta\tau$ used to approximate the continuous evolution of the system introduces a systematic time step error~\cite{AssaadReview}. Specifically, in this work, the symmetric Trotter decomposition leads to an error proportional to $\Delta \tau^2$, while the discrete HS transformation introduces an additional systematic error of order $\mathcal{O}(\Delta \tau^3)$. Focusing only on the leading-order error, the expectation value of an observable $O$ computed with a finite $\Delta\tau$ deviates from its exact value $\braket{O}_{\mathrm{exact}}$ as:
\begin{equation}
	\braket{O}_{\Delta\tau} = \braket{O}_{\mathrm{exact}} + C_O\Delta\tau^2,
	\label{eq:deltatau}
\end{equation}
where $C_O$ is an observable-dependent coefficient that characterizes the leading-order time step error.

\begin{figure}[htbp]
	\centering
	\includegraphics[width=0.6\linewidth]{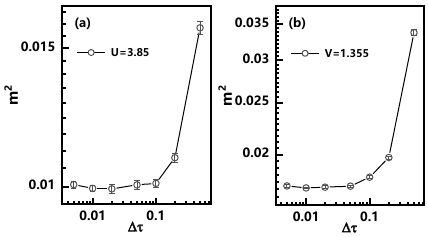}
	\vskip-3mm
	\caption{Log-log plots of order parameter $m^2$ as a function of time step $\Delta\tau$. \textbf{(a)} Results for the honeycomb Hubbard model, driven from DSM phase $U=0$ to the critical point $U_c=3.85$, using 1920 samples with tuning parameters $L=12$, $R=0.5$. \textbf{(b)} Results for the $t-V$ model, driven from DSM phase $V=0$ to the critical point $V_c=1.355$, using 1200 samples under otherwise identical conditions.}
	\label{time step error}
\end{figure}

Here, we numerically demonstrate that our choice of $\Delta\tau=0.05$ is sufficiently small, ensuring that its impact on our results is negligible. As shown in Fig.~\ref{time step error}, with a sample count similar to that in the main text, the curves flatten for $\Delta\tau \leq 0.05$, indicating that the time step error is negligible and the results closely approximate the $\Delta\tau \to 0$ limit. In this regime, the fluctuations in the numerical results are dominated by statistical noise from Monte Carlo sampling rather than systematic errors from the time discretization.


\begin{thebibliography}{75}%
    \makeatletter
    \providecommand \@ifxundefined [1]{%
     \@ifx{#1\undefined}
    }%
    \providecommand \@ifnum [1]{%
     \ifnum #1\expandafter \@firstoftwo
     \else \expandafter \@secondoftwo
     \fi
    }%
    \providecommand \@ifx [1]{%
     \ifx #1\expandafter \@firstoftwo
     \else \expandafter \@secondoftwo
     \fi
    }%
    \providecommand \natexlab [1]{#1}%
    \providecommand \enquote  [1]{``#1''}%
    \providecommand \bibnamefont  [1]{#1}%
    \providecommand \bibfnamefont [1]{#1}%
    \providecommand \citenamefont [1]{#1}%
    \providecommand \href@noop [0]{\@secondoftwo}%
    \providecommand \href [0]{\begingroup \@sanitize@url \@href}%
    \providecommand \@href[1]{\@@startlink{#1}\@@href}%
    \providecommand \@@href[1]{\endgroup#1\@@endlink}%
    \providecommand \@sanitize@url [0]{\catcode `\\12\catcode `\$12\catcode `\&12\catcode `\#12\catcode `\^12\catcode `\_12\catcode `\%12\relax}%
    \providecommand \@@startlink[1]{}%
    \providecommand \@@endlink[0]{}%
    \providecommand \url  [0]{\begingroup\@sanitize@url \@url }%
    \providecommand \@url [1]{\endgroup\@href {#1}{\urlprefix }}%
    \providecommand \urlprefix  [0]{URL }%
    \providecommand \Eprint [0]{\href }%
    \providecommand \doibase [0]{https://doi.org/}%
    \providecommand \selectlanguage [0]{\@gobble}%
    \providecommand \bibinfo  [0]{\@secondoftwo}%
    \providecommand \bibfield  [0]{\@secondoftwo}%
    \providecommand \translation [1]{[#1]}%
    \providecommand \BibitemOpen [0]{}%
    \providecommand \bibitemStop [0]{}%
    \providecommand \bibitemNoStop [0]{.\EOS\space}%
    \providecommand \EOS [0]{\spacefactor3000\relax}%
    \providecommand \BibitemShut  [1]{\csname bibitem#1\endcsname}%
    \let\auto@bib@innerbib\@empty
    \bibitem [{\citenamefont {Dziarmaga}(2010)}]{Dziarmaga2010review}%
      \BibitemOpen
      \bibfield  {author} {\bibinfo {author} {\bibfnamefont {J.}~\bibnamefont {Dziarmaga}},\ }\href {https://doi.org/10.1080/00018732.2010.514702} {\bibfield  {journal} {\bibinfo  {journal} {Advances in Physics}\ }\textbf {\bibinfo {volume} {59}},\ \bibinfo {pages} {1063} (\bibinfo {year} {2010})}\BibitemShut {NoStop}%
    \bibitem [{\citenamefont {Polkovnikov}\ \emph {et~al.}(2011)\citenamefont {Polkovnikov}, \citenamefont {Sengupta}, \citenamefont {Silva},\ and\ \citenamefont {Vengalattore}}]{Polkovnikov2011rmp}%
      \BibitemOpen
      \bibfield  {author} {\bibinfo {author} {\bibfnamefont {A.}~\bibnamefont {Polkovnikov}}, \bibinfo {author} {\bibfnamefont {K.}~\bibnamefont {Sengupta}}, \bibinfo {author} {\bibfnamefont {A.}~\bibnamefont {Silva}},\ and\ \bibinfo {author} {\bibfnamefont {M.}~\bibnamefont {Vengalattore}},\ }\href {https://doi.org/10.1103/RevModPhys.83.863} {\bibfield  {journal} {\bibinfo  {journal} {Rev. Mod. Phys.}\ }\textbf {\bibinfo {volume} {83}},\ \bibinfo {pages} {863} (\bibinfo {year} {2011})}\BibitemShut {NoStop}%
    \bibitem [{\citenamefont {Kibble}(1976)}]{Kibble1976}%
      \BibitemOpen
      \bibfield  {author} {\bibinfo {author} {\bibfnamefont {T.~W.~B.}\ \bibnamefont {Kibble}},\ }\href {https://doi.org/10.1088/0305-4470/9/8/029} {\bibfield  {journal} {\bibinfo  {journal} {Journal of Physics A: Mathematical and General}\ }\textbf {\bibinfo {volume} {9}},\ \bibinfo {pages} {1387} (\bibinfo {year} {1976})}\BibitemShut {NoStop}%
    \bibitem [{\citenamefont {Zurek}(1985)}]{Zurek1985}%
      \BibitemOpen
      \bibfield  {author} {\bibinfo {author} {\bibfnamefont {W.~H.}\ \bibnamefont {Zurek}},\ }\href {https://doi.org/10.1038/317505a0} {\bibfield  {journal} {\bibinfo  {journal} {Nature}\ }\textbf {\bibinfo {volume} {317}},\ \bibinfo {pages} {505} (\bibinfo {year} {1985})}\BibitemShut {NoStop}%
    \bibitem [{\citenamefont {Zurek}\ \emph {et~al.}(2005)\citenamefont {Zurek}, \citenamefont {Dorner},\ and\ \citenamefont {Zoller}}]{Zoller2005prl}%
      \BibitemOpen
      \bibfield  {author} {\bibinfo {author} {\bibfnamefont {W.~H.}\ \bibnamefont {Zurek}}, \bibinfo {author} {\bibfnamefont {U.}~\bibnamefont {Dorner}},\ and\ \bibinfo {author} {\bibfnamefont {P.}~\bibnamefont {Zoller}},\ }\href {https://doi.org/10.1103/PhysRevLett.95.105701} {\bibfield  {journal} {\bibinfo  {journal} {Phys. Rev. Lett.}\ }\textbf {\bibinfo {volume} {95}},\ \bibinfo {pages} {105701} (\bibinfo {year} {2005})}\BibitemShut {NoStop}%
    \bibitem [{\citenamefont {Dziarmaga}(2005)}]{Dziarmaga2005prl}%
      \BibitemOpen
      \bibfield  {author} {\bibinfo {author} {\bibfnamefont {J.}~\bibnamefont {Dziarmaga}},\ }\href {https://doi.org/10.1103/PhysRevLett.95.245701} {\bibfield  {journal} {\bibinfo  {journal} {Phys. Rev. Lett.}\ }\textbf {\bibinfo {volume} {95}},\ \bibinfo {pages} {245701} (\bibinfo {year} {2005})}\BibitemShut {NoStop}%
    \bibitem [{\citenamefont {Polkovnikov}(2005)}]{PhysRevB.72.161201}%
      \BibitemOpen
      \bibfield  {author} {\bibinfo {author} {\bibfnamefont {A.}~\bibnamefont {Polkovnikov}},\ }\href {https://doi.org/10.1103/PhysRevB.72.161201} {\bibfield  {journal} {\bibinfo  {journal} {Phys. Rev. B}\ }\textbf {\bibinfo {volume} {72}},\ \bibinfo {pages} {161201} (\bibinfo {year} {2005})}\BibitemShut {NoStop}%
    \bibitem [{\citenamefont {Du}\ \emph {et~al.}(2023)\citenamefont {Du}, \citenamefont {Fang}, \citenamefont {Won}, \citenamefont {De}, \citenamefont {Huang}, \citenamefont {Xu}, \citenamefont {You}, \citenamefont {Gómez-Ruiz}, \citenamefont {del Campo},\ and\ \citenamefont {Cheong}}]{Du2023}%
      \BibitemOpen
      \bibfield  {author} {\bibinfo {author} {\bibfnamefont {K.}~\bibnamefont {Du}}, \bibinfo {author} {\bibfnamefont {X.}~\bibnamefont {Fang}}, \bibinfo {author} {\bibfnamefont {C.}~\bibnamefont {Won}}, \bibinfo {author} {\bibfnamefont {C.}~\bibnamefont {De}}, \bibinfo {author} {\bibfnamefont {F.-T.}\ \bibnamefont {Huang}}, \bibinfo {author} {\bibfnamefont {W.}~\bibnamefont {Xu}}, \bibinfo {author} {\bibfnamefont {H.}~\bibnamefont {You}}, \bibinfo {author} {\bibfnamefont {F.~J.}\ \bibnamefont {Gómez-Ruiz}}, \bibinfo {author} {\bibfnamefont {A.}~\bibnamefont {del Campo}},\ and\ \bibinfo {author} {\bibfnamefont {S.-W.}\ \bibnamefont {Cheong}},\ }\href {https://doi.org/10.1038/s41567-023-02112-5} {\bibfield  {journal} {\bibinfo  {journal} {Nature Physics}\ }\textbf {\bibinfo {volume} {19}},\ \bibinfo {pages} {1495} (\bibinfo {year} {2023})}\BibitemShut {NoStop}%
    \bibitem [{\citenamefont {Ko}\ \emph {et~al.}(2019)\citenamefont {Ko}, \citenamefont {Park},\ and\ \citenamefont {Shin}}]{Ko2019}%
      \BibitemOpen
      \bibfield  {author} {\bibinfo {author} {\bibfnamefont {B.}~\bibnamefont {Ko}}, \bibinfo {author} {\bibfnamefont {J.~W.}\ \bibnamefont {Park}},\ and\ \bibinfo {author} {\bibfnamefont {Y.}~\bibnamefont {Shin}},\ }\href {https://doi.org/10.1038/s41567-019-0650-1} {\bibfield  {journal} {\bibinfo  {journal} {Nature Physics}\ }\textbf {\bibinfo {volume} {15}},\ \bibinfo {pages} {1227} (\bibinfo {year} {2019})}\BibitemShut {NoStop}%
    \bibitem [{\citenamefont {Maegochi}\ \emph {et~al.}(2022)\citenamefont {Maegochi}, \citenamefont {Ienaga},\ and\ \citenamefont {Okuma}}]{PhysRevLett.129.227001}%
      \BibitemOpen
      \bibfield  {author} {\bibinfo {author} {\bibfnamefont {S.}~\bibnamefont {Maegochi}}, \bibinfo {author} {\bibfnamefont {K.}~\bibnamefont {Ienaga}},\ and\ \bibinfo {author} {\bibfnamefont {S.}~\bibnamefont {Okuma}},\ }\href {https://doi.org/10.1103/PhysRevLett.129.227001} {\bibfield  {journal} {\bibinfo  {journal} {Phys. Rev. Lett.}\ }\textbf {\bibinfo {volume} {129}},\ \bibinfo {pages} {227001} (\bibinfo {year} {2022})}\BibitemShut {NoStop}%
    \bibitem [{\citenamefont {Keesling}\ \emph {et~al.}(2019)\citenamefont {Keesling}, \citenamefont {Omran}, \citenamefont {Levine}, \citenamefont {Bernien}, \citenamefont {Pichler}, \citenamefont {Choi}, \citenamefont {Samajdar}, \citenamefont {Schwartz}, \citenamefont {Silvi}, \citenamefont {Sachdev}, \citenamefont {Zoller}, \citenamefont {Endres}, \citenamefont {Greiner}, \citenamefont {Vuleti{\'{c}}},\ and\ \citenamefont {Lukin}}]{Keesling2019}%
      \BibitemOpen
      \bibfield  {author} {\bibinfo {author} {\bibfnamefont {A.}~\bibnamefont {Keesling}}, \bibinfo {author} {\bibfnamefont {A.}~\bibnamefont {Omran}}, \bibinfo {author} {\bibfnamefont {H.}~\bibnamefont {Levine}}, \bibinfo {author} {\bibfnamefont {H.}~\bibnamefont {Bernien}}, \bibinfo {author} {\bibfnamefont {H.}~\bibnamefont {Pichler}}, \bibinfo {author} {\bibfnamefont {S.}~\bibnamefont {Choi}}, \bibinfo {author} {\bibfnamefont {R.}~\bibnamefont {Samajdar}}, \bibinfo {author} {\bibfnamefont {S.}~\bibnamefont {Schwartz}}, \bibinfo {author} {\bibfnamefont {P.}~\bibnamefont {Silvi}}, \bibinfo {author} {\bibfnamefont {S.}~\bibnamefont {Sachdev}}, \bibinfo {author} {\bibfnamefont {P.}~\bibnamefont {Zoller}}, \bibinfo {author} {\bibfnamefont {M.}~\bibnamefont {Endres}}, \bibinfo {author} {\bibfnamefont {M.}~\bibnamefont {Greiner}}, \bibinfo {author} {\bibfnamefont {V.}~\bibnamefont {Vuleti{\'{c}}}},\ and\ \bibinfo {author} {\bibfnamefont {M.~D.}\ \bibnamefont {Lukin}},\ }\href
      {https://doi.org/10.1038/s41586-019-1070-1} {\bibfield  {journal} {\bibinfo  {journal} {Nature}\ }\textbf {\bibinfo {volume} {568}},\ \bibinfo {pages} {207} (\bibinfo {year} {2019})}\BibitemShut {NoStop}%
    \bibitem [{\citenamefont {Ebadi}\ \emph {et~al.}(2021)\citenamefont {Ebadi}, \citenamefont {Wang}, \citenamefont {Levine}, \citenamefont {Keesling}, \citenamefont {Semeghini}, \citenamefont {Omran}, \citenamefont {Bluvstein}, \citenamefont {Samajdar}, \citenamefont {Pichler}, \citenamefont {Ho}, \citenamefont {Choi}, \citenamefont {Sachdev}, \citenamefont {Greiner}, \citenamefont {Vuleti{\'{c}}},\ and\ \citenamefont {Lukin}}]{Ebadi2021}%
      \BibitemOpen
      \bibfield  {author} {\bibinfo {author} {\bibfnamefont {S.}~\bibnamefont {Ebadi}}, \bibinfo {author} {\bibfnamefont {T.~T.}\ \bibnamefont {Wang}}, \bibinfo {author} {\bibfnamefont {H.}~\bibnamefont {Levine}}, \bibinfo {author} {\bibfnamefont {A.}~\bibnamefont {Keesling}}, \bibinfo {author} {\bibfnamefont {G.}~\bibnamefont {Semeghini}}, \bibinfo {author} {\bibfnamefont {A.}~\bibnamefont {Omran}}, \bibinfo {author} {\bibfnamefont {D.}~\bibnamefont {Bluvstein}}, \bibinfo {author} {\bibfnamefont {R.}~\bibnamefont {Samajdar}}, \bibinfo {author} {\bibfnamefont {H.}~\bibnamefont {Pichler}}, \bibinfo {author} {\bibfnamefont {W.~W.}\ \bibnamefont {Ho}}, \bibinfo {author} {\bibfnamefont {S.}~\bibnamefont {Choi}}, \bibinfo {author} {\bibfnamefont {S.}~\bibnamefont {Sachdev}}, \bibinfo {author} {\bibfnamefont {M.}~\bibnamefont {Greiner}}, \bibinfo {author} {\bibfnamefont {V.}~\bibnamefont {Vuleti{\'{c}}}},\ and\ \bibinfo {author} {\bibfnamefont {M.~D.}\ \bibnamefont {Lukin}},\ }\href
      {https://doi.org/10.1038/s41586-021-03582-4} {\bibfield  {journal} {\bibinfo  {journal} {Nature}\ }\textbf {\bibinfo {volume} {595}},\ \bibinfo {pages} {227} (\bibinfo {year} {2021})}\BibitemShut {NoStop}%
    \bibitem [{\citenamefont {Qiu}\ \emph {et~al.}(2020)\citenamefont {Qiu}, \citenamefont {Liang}, \citenamefont {Yang}, \citenamefont {Yang}, \citenamefont {Tian}, \citenamefont {Xu},\ and\ \citenamefont {Duan}}]{sciadv.aba7292}%
      \BibitemOpen
      \bibfield  {author} {\bibinfo {author} {\bibfnamefont {L.-Y.}\ \bibnamefont {Qiu}}, \bibinfo {author} {\bibfnamefont {H.-Y.}\ \bibnamefont {Liang}}, \bibinfo {author} {\bibfnamefont {Y.-B.}\ \bibnamefont {Yang}}, \bibinfo {author} {\bibfnamefont {H.-X.}\ \bibnamefont {Yang}}, \bibinfo {author} {\bibfnamefont {T.}~\bibnamefont {Tian}}, \bibinfo {author} {\bibfnamefont {Y.}~\bibnamefont {Xu}},\ and\ \bibinfo {author} {\bibfnamefont {L.-M.}\ \bibnamefont {Duan}},\ }\href {https://doi.org/10.1126/sciadv.aba7292} {\bibfield  {journal} {\bibinfo  {journal} {Science Advances}\ }\textbf {\bibinfo {volume} {6}},\ \bibinfo {pages} {eaba7292} (\bibinfo {year} {2020})},\ \Eprint {https://arxiv.org/abs/https://www.science.org/doi/pdf/10.1126/sciadv.aba7292} {https://www.science.org/doi/pdf/10.1126/sciadv.aba7292} \BibitemShut {NoStop}%
    \bibitem [{\citenamefont {Ebadi}\ \emph {et~al.}(2022)\citenamefont {Ebadi}, \citenamefont {Keesling}, \citenamefont {Cain}, \citenamefont {Wang}, \citenamefont {Levine}, \citenamefont {Bluvstein}, \citenamefont {Semeghini}, \citenamefont {Omran}, \citenamefont {Liu}, \citenamefont {Samajdar}, \citenamefont {Luo}, \citenamefont {Nash}, \citenamefont {Gao}, \citenamefont {Barak}, \citenamefont {Farhi}, \citenamefont {Sachdev}, \citenamefont {Gemelke}, \citenamefont {Zhou}, \citenamefont {Choi}, \citenamefont {Pichler}, \citenamefont {Wang}, \citenamefont {Greiner}, \citenamefont {Vuletić},\ and\ \citenamefont {Lukin}}]{science.abo6587}%
      \BibitemOpen
      \bibfield  {author} {\bibinfo {author} {\bibfnamefont {S.}~\bibnamefont {Ebadi}}, \bibinfo {author} {\bibfnamefont {A.}~\bibnamefont {Keesling}}, \bibinfo {author} {\bibfnamefont {M.}~\bibnamefont {Cain}}, \bibinfo {author} {\bibfnamefont {T.~T.}\ \bibnamefont {Wang}}, \bibinfo {author} {\bibfnamefont {H.}~\bibnamefont {Levine}}, \bibinfo {author} {\bibfnamefont {D.}~\bibnamefont {Bluvstein}}, \bibinfo {author} {\bibfnamefont {G.}~\bibnamefont {Semeghini}}, \bibinfo {author} {\bibfnamefont {A.}~\bibnamefont {Omran}}, \bibinfo {author} {\bibfnamefont {J.-G.}\ \bibnamefont {Liu}}, \bibinfo {author} {\bibfnamefont {R.}~\bibnamefont {Samajdar}}, \bibinfo {author} {\bibfnamefont {X.-Z.}\ \bibnamefont {Luo}}, \bibinfo {author} {\bibfnamefont {B.}~\bibnamefont {Nash}}, \bibinfo {author} {\bibfnamefont {X.}~\bibnamefont {Gao}}, \bibinfo {author} {\bibfnamefont {B.}~\bibnamefont {Barak}}, \bibinfo {author} {\bibfnamefont {E.}~\bibnamefont {Farhi}}, \bibinfo {author} {\bibfnamefont {S.}~\bibnamefont {Sachdev}},
      \bibinfo {author} {\bibfnamefont {N.}~\bibnamefont {Gemelke}}, \bibinfo {author} {\bibfnamefont {L.}~\bibnamefont {Zhou}}, \bibinfo {author} {\bibfnamefont {S.}~\bibnamefont {Choi}}, \bibinfo {author} {\bibfnamefont {H.}~\bibnamefont {Pichler}}, \bibinfo {author} {\bibfnamefont {S.-T.}\ \bibnamefont {Wang}}, \bibinfo {author} {\bibfnamefont {M.}~\bibnamefont {Greiner}}, \bibinfo {author} {\bibfnamefont {V.}~\bibnamefont {Vuletić}},\ and\ \bibinfo {author} {\bibfnamefont {M.~D.}\ \bibnamefont {Lukin}},\ }\href {https://doi.org/10.1126/science.abo6587} {\bibfield  {journal} {\bibinfo  {journal} {Science}\ }\textbf {\bibinfo {volume} {376}},\ \bibinfo {pages} {1209} (\bibinfo {year} {2022})},\ \Eprint {https://arxiv.org/abs/https://www.science.org/doi/pdf/10.1126/science.abo6587} {https://www.science.org/doi/pdf/10.1126/science.abo6587} \BibitemShut {NoStop}%
    \bibitem [{\citenamefont {Sunami}\ \emph {et~al.}(2023)\citenamefont {Sunami}, \citenamefont {Singh}, \citenamefont {Garrick}, \citenamefont {Beregi}, \citenamefont {Barker}, \citenamefont {Luksch}, \citenamefont {Bentine}, \citenamefont {Mathey},\ and\ \citenamefont {Foot}}]{science.abq6753}%
      \BibitemOpen
      \bibfield  {author} {\bibinfo {author} {\bibfnamefont {S.}~\bibnamefont {Sunami}}, \bibinfo {author} {\bibfnamefont {V.~P.}\ \bibnamefont {Singh}}, \bibinfo {author} {\bibfnamefont {D.}~\bibnamefont {Garrick}}, \bibinfo {author} {\bibfnamefont {A.}~\bibnamefont {Beregi}}, \bibinfo {author} {\bibfnamefont {A.~J.}\ \bibnamefont {Barker}}, \bibinfo {author} {\bibfnamefont {K.}~\bibnamefont {Luksch}}, \bibinfo {author} {\bibfnamefont {E.}~\bibnamefont {Bentine}}, \bibinfo {author} {\bibfnamefont {L.}~\bibnamefont {Mathey}},\ and\ \bibinfo {author} {\bibfnamefont {C.~J.}\ \bibnamefont {Foot}},\ }\href {https://doi.org/10.1126/science.abq6753} {\bibfield  {journal} {\bibinfo  {journal} {Science}\ }\textbf {\bibinfo {volume} {382}},\ \bibinfo {pages} {443} (\bibinfo {year} {2023})},\ \Eprint {https://arxiv.org/abs/https://www.science.org/doi/pdf/10.1126/science.abq6753} {https://www.science.org/doi/pdf/10.1126/science.abq6753} \BibitemShut {NoStop}%
    \bibitem [{\citenamefont {Li}\ \emph {et~al.}(2023)\citenamefont {Li}, \citenamefont {Wu}, \citenamefont {Mei}, \citenamefont {Yao}, \citenamefont {Lian}, \citenamefont {Cai}, \citenamefont {Wang}, \citenamefont {Qi}, \citenamefont {Yao}, \citenamefont {He}, \citenamefont {Zhou},\ and\ \citenamefont {Duan}}]{PRXQuantum}%
      \BibitemOpen
      \bibfield  {author} {\bibinfo {author} {\bibfnamefont {B.-W.}\ \bibnamefont {Li}}, \bibinfo {author} {\bibfnamefont {Y.-K.}\ \bibnamefont {Wu}}, \bibinfo {author} {\bibfnamefont {Q.-X.}\ \bibnamefont {Mei}}, \bibinfo {author} {\bibfnamefont {R.}~\bibnamefont {Yao}}, \bibinfo {author} {\bibfnamefont {W.-Q.}\ \bibnamefont {Lian}}, \bibinfo {author} {\bibfnamefont {M.-L.}\ \bibnamefont {Cai}}, \bibinfo {author} {\bibfnamefont {Y.}~\bibnamefont {Wang}}, \bibinfo {author} {\bibfnamefont {B.-X.}\ \bibnamefont {Qi}}, \bibinfo {author} {\bibfnamefont {L.}~\bibnamefont {Yao}}, \bibinfo {author} {\bibfnamefont {L.}~\bibnamefont {He}}, \bibinfo {author} {\bibfnamefont {Z.-C.}\ \bibnamefont {Zhou}},\ and\ \bibinfo {author} {\bibfnamefont {L.-M.}\ \bibnamefont {Duan}},\ }\href {https://doi.org/10.1103/PRXQuantum.4.010302} {\bibfield  {journal} {\bibinfo  {journal} {PRX Quantum}\ }\textbf {\bibinfo {volume} {4}},\ \bibinfo {pages} {010302} (\bibinfo {year} {2023})}\BibitemShut {NoStop}%
    \bibitem [{\citenamefont {Zhong}\ and\ \citenamefont {Xu}(2005)}]{Zhifangxu2005prb}%
      \BibitemOpen
      \bibfield  {author} {\bibinfo {author} {\bibfnamefont {F.}~\bibnamefont {Zhong}}\ and\ \bibinfo {author} {\bibfnamefont {Z.}~\bibnamefont {Xu}},\ }\href {https://doi.org/10.1103/PhysRevB.71.132402} {\bibfield  {journal} {\bibinfo  {journal} {Phys. Rev. B}\ }\textbf {\bibinfo {volume} {71}},\ \bibinfo {pages} {132402} (\bibinfo {year} {2005})}\BibitemShut {NoStop}%
    \bibitem [{\citenamefont {Deng}\ \emph {et~al.}(2009)\citenamefont {Deng}, \citenamefont {Ortiz},\ and\ \citenamefont {Viola}}]{Deng2008epl}%
      \BibitemOpen
      \bibfield  {author} {\bibinfo {author} {\bibfnamefont {S.}~\bibnamefont {Deng}}, \bibinfo {author} {\bibfnamefont {G.}~\bibnamefont {Ortiz}},\ and\ \bibinfo {author} {\bibfnamefont {L.}~\bibnamefont {Viola}},\ }\href {https://doi.org/10.1209/0295-5075/84/67008} {\bibfield  {journal} {\bibinfo  {journal} {Europhysics Letters}\ }\textbf {\bibinfo {volume} {84}},\ \bibinfo {pages} {67008} (\bibinfo {year} {2009})}\BibitemShut {NoStop}%
    \bibitem [{\citenamefont {Chandran}\ \emph {et~al.}(2012)\citenamefont {Chandran}, \citenamefont {Erez}, \citenamefont {Gubser},\ and\ \citenamefont {Sondhi}}]{Chandran2012prb}%
      \BibitemOpen
      \bibfield  {author} {\bibinfo {author} {\bibfnamefont {A.}~\bibnamefont {Chandran}}, \bibinfo {author} {\bibfnamefont {A.}~\bibnamefont {Erez}}, \bibinfo {author} {\bibfnamefont {S.~S.}\ \bibnamefont {Gubser}},\ and\ \bibinfo {author} {\bibfnamefont {S.~L.}\ \bibnamefont {Sondhi}},\ }\href {https://doi.org/10.1103/PhysRevB.86.064304} {\bibfield  {journal} {\bibinfo  {journal} {Phys. Rev. B}\ }\textbf {\bibinfo {volume} {86}},\ \bibinfo {pages} {064304} (\bibinfo {year} {2012})}\BibitemShut {NoStop}%
    \bibitem [{\citenamefont {Clark}\ \emph {et~al.}(2016)\citenamefont {Clark}, \citenamefont {Feng},\ and\ \citenamefont {Chin}}]{Clark2016science}%
      \BibitemOpen
      \bibfield  {author} {\bibinfo {author} {\bibfnamefont {L.~W.}\ \bibnamefont {Clark}}, \bibinfo {author} {\bibfnamefont {L.}~\bibnamefont {Feng}},\ and\ \bibinfo {author} {\bibfnamefont {C.}~\bibnamefont {Chin}},\ }\href {https://doi.org/10.1126/science.aaf9657} {\bibfield  {journal} {\bibinfo  {journal} {Science}\ }\textbf {\bibinfo {volume} {354}},\ \bibinfo {pages} {606} (\bibinfo {year} {2016})},\ \Eprint {https://arxiv.org/abs/https://www.science.org/doi/pdf/10.1126/science.aaf9657} {https://www.science.org/doi/pdf/10.1126/science.aaf9657} \BibitemShut {NoStop}%
    \bibitem [{\citenamefont {Kolodrubetz}\ \emph {et~al.}(2012)\citenamefont {Kolodrubetz}, \citenamefont {Clark},\ and\ \citenamefont {Huse}}]{Huse2012prl}%
      \BibitemOpen
      \bibfield  {author} {\bibinfo {author} {\bibfnamefont {M.}~\bibnamefont {Kolodrubetz}}, \bibinfo {author} {\bibfnamefont {B.~K.}\ \bibnamefont {Clark}},\ and\ \bibinfo {author} {\bibfnamefont {D.~A.}\ \bibnamefont {Huse}},\ }\href {https://doi.org/10.1103/PhysRevLett.109.015701} {\bibfield  {journal} {\bibinfo  {journal} {Phys. Rev. Lett.}\ }\textbf {\bibinfo {volume} {109}},\ \bibinfo {pages} {015701} (\bibinfo {year} {2012})}\BibitemShut {NoStop}%
    \bibitem [{\citenamefont {Gong}\ \emph {et~al.}(2010)\citenamefont {Gong}, \citenamefont {Zhong}, \citenamefont {Huang},\ and\ \citenamefont {Fan}}]{Gong2010njp}%
      \BibitemOpen
      \bibfield  {author} {\bibinfo {author} {\bibfnamefont {S.}~\bibnamefont {Gong}}, \bibinfo {author} {\bibfnamefont {F.}~\bibnamefont {Zhong}}, \bibinfo {author} {\bibfnamefont {X.}~\bibnamefont {Huang}},\ and\ \bibinfo {author} {\bibfnamefont {S.}~\bibnamefont {Fan}},\ }\href {https://doi.org/10.1088/1367-2630/12/4/043036} {\bibfield  {journal} {\bibinfo  {journal} {New Journal of Physics}\ }\textbf {\bibinfo {volume} {12}},\ \bibinfo {pages} {043036} (\bibinfo {year} {2010})}\BibitemShut {NoStop}%
    \bibitem [{\citenamefont {Feng}\ \emph {et~al.}(2016)\citenamefont {Feng}, \citenamefont {Yin},\ and\ \citenamefont {Zhong}}]{Feng2016prb}%
      \BibitemOpen
      \bibfield  {author} {\bibinfo {author} {\bibfnamefont {B.}~\bibnamefont {Feng}}, \bibinfo {author} {\bibfnamefont {S.}~\bibnamefont {Yin}},\ and\ \bibinfo {author} {\bibfnamefont {F.}~\bibnamefont {Zhong}},\ }\href {https://doi.org/10.1103/PhysRevB.94.144103} {\bibfield  {journal} {\bibinfo  {journal} {Phys. Rev. B}\ }\textbf {\bibinfo {volume} {94}},\ \bibinfo {pages} {144103} (\bibinfo {year} {2016})}\BibitemShut {NoStop}%
    \bibitem [{\citenamefont {Huang}\ \emph {et~al.}(2014)\citenamefont {Huang}, \citenamefont {Yin}, \citenamefont {Feng},\ and\ \citenamefont {Zhong}}]{huangyy2014prb}%
      \BibitemOpen
      \bibfield  {author} {\bibinfo {author} {\bibfnamefont {Y.}~\bibnamefont {Huang}}, \bibinfo {author} {\bibfnamefont {S.}~\bibnamefont {Yin}}, \bibinfo {author} {\bibfnamefont {B.}~\bibnamefont {Feng}},\ and\ \bibinfo {author} {\bibfnamefont {F.}~\bibnamefont {Zhong}},\ }\href {https://doi.org/10.1103/PhysRevB.90.134108} {\bibfield  {journal} {\bibinfo  {journal} {Phys. Rev. B}\ }\textbf {\bibinfo {volume} {90}},\ \bibinfo {pages} {134108} (\bibinfo {year} {2014})}\BibitemShut {NoStop}%
    \bibitem [{\citenamefont {Liu}\ \emph {et~al.}(2014)\citenamefont {Liu}, \citenamefont {Polkovnikov},\ and\ \citenamefont {Sandvik}}]{Liuchengwei2014prb}%
      \BibitemOpen
      \bibfield  {author} {\bibinfo {author} {\bibfnamefont {C.-W.}\ \bibnamefont {Liu}}, \bibinfo {author} {\bibfnamefont {A.}~\bibnamefont {Polkovnikov}},\ and\ \bibinfo {author} {\bibfnamefont {A.~W.}\ \bibnamefont {Sandvik}},\ }\href {https://doi.org/10.1103/PhysRevB.89.054307} {\bibfield  {journal} {\bibinfo  {journal} {Phys. Rev. B}\ }\textbf {\bibinfo {volume} {89}},\ \bibinfo {pages} {054307} (\bibinfo {year} {2014})}\BibitemShut {NoStop}%
    \bibitem [{\citenamefont {Yin}\ \emph {et~al.}(2014)\citenamefont {Yin}, \citenamefont {Mai},\ and\ \citenamefont {Zhong}}]{Yin2014prb}%
      \BibitemOpen
      \bibfield  {author} {\bibinfo {author} {\bibfnamefont {S.}~\bibnamefont {Yin}}, \bibinfo {author} {\bibfnamefont {P.}~\bibnamefont {Mai}},\ and\ \bibinfo {author} {\bibfnamefont {F.}~\bibnamefont {Zhong}},\ }\href {https://doi.org/10.1103/PhysRevB.89.094108} {\bibfield  {journal} {\bibinfo  {journal} {Phys. Rev. B}\ }\textbf {\bibinfo {volume} {89}},\ \bibinfo {pages} {094108} (\bibinfo {year} {2014})}\BibitemShut {NoStop}%
    \bibitem [{\citenamefont {Liu}\ \emph {et~al.}(2015)\citenamefont {Liu}, \citenamefont {Polkovnikov},\ and\ \citenamefont {Sandvik}}]{Sandvik2015prl}%
      \BibitemOpen
      \bibfield  {author} {\bibinfo {author} {\bibfnamefont {C.-W.}\ \bibnamefont {Liu}}, \bibinfo {author} {\bibfnamefont {A.}~\bibnamefont {Polkovnikov}},\ and\ \bibinfo {author} {\bibfnamefont {A.~W.}\ \bibnamefont {Sandvik}},\ }\href {https://doi.org/10.1103/PhysRevLett.114.147203} {\bibfield  {journal} {\bibinfo  {journal} {Phys. Rev. Lett.}\ }\textbf {\bibinfo {volume} {114}},\ \bibinfo {pages} {147203} (\bibinfo {year} {2015})}\BibitemShut {NoStop}%
    \bibitem [{\citenamefont {King}\ \emph {et~al.}(2023)\citenamefont {King}, \citenamefont {Raymond}, \citenamefont {Lanting}, \citenamefont {Harris}, \citenamefont {Zucca}, \citenamefont {Altomare}, \citenamefont {Berkley}, \citenamefont {Boothby}, \citenamefont {Ejtemaee}, \citenamefont {Enderud}, \citenamefont {Hoskinson}, \citenamefont {Huang}, \citenamefont {Ladizinsky}, \citenamefont {MacDonald}, \citenamefont {Marsden}, \citenamefont {Molavi}, \citenamefont {Oh}, \citenamefont {Poulin-Lamarre}, \citenamefont {Reis}, \citenamefont {Rich}, \citenamefont {Sato}, \citenamefont {Tsai}, \citenamefont {Volkmann}, \citenamefont {Whittaker}, \citenamefont {Yao}, \citenamefont {Sandvik},\ and\ \citenamefont {Amin}}]{king2023nature}%
      \BibitemOpen
      \bibfield  {author} {\bibinfo {author} {\bibfnamefont {A.~D.}\ \bibnamefont {King}}, \bibinfo {author} {\bibfnamefont {J.}~\bibnamefont {Raymond}}, \bibinfo {author} {\bibfnamefont {T.}~\bibnamefont {Lanting}}, \bibinfo {author} {\bibfnamefont {R.}~\bibnamefont {Harris}}, \bibinfo {author} {\bibfnamefont {A.}~\bibnamefont {Zucca}}, \bibinfo {author} {\bibfnamefont {F.}~\bibnamefont {Altomare}}, \bibinfo {author} {\bibfnamefont {A.~J.}\ \bibnamefont {Berkley}}, \bibinfo {author} {\bibfnamefont {K.}~\bibnamefont {Boothby}}, \bibinfo {author} {\bibfnamefont {S.}~\bibnamefont {Ejtemaee}}, \bibinfo {author} {\bibfnamefont {C.}~\bibnamefont {Enderud}}, \bibinfo {author} {\bibfnamefont {E.}~\bibnamefont {Hoskinson}}, \bibinfo {author} {\bibfnamefont {S.}~\bibnamefont {Huang}}, \bibinfo {author} {\bibfnamefont {E.}~\bibnamefont {Ladizinsky}}, \bibinfo {author} {\bibfnamefont {A.~J.~R.}\ \bibnamefont {MacDonald}}, \bibinfo {author} {\bibfnamefont {G.}~\bibnamefont {Marsden}}, \bibinfo {author} {\bibfnamefont
      {R.}~\bibnamefont {Molavi}}, \bibinfo {author} {\bibfnamefont {T.}~\bibnamefont {Oh}}, \bibinfo {author} {\bibfnamefont {G.}~\bibnamefont {Poulin-Lamarre}}, \bibinfo {author} {\bibfnamefont {M.}~\bibnamefont {Reis}}, \bibinfo {author} {\bibfnamefont {C.}~\bibnamefont {Rich}}, \bibinfo {author} {\bibfnamefont {Y.}~\bibnamefont {Sato}}, \bibinfo {author} {\bibfnamefont {N.}~\bibnamefont {Tsai}}, \bibinfo {author} {\bibfnamefont {M.}~\bibnamefont {Volkmann}}, \bibinfo {author} {\bibfnamefont {J.~D.}\ \bibnamefont {Whittaker}}, \bibinfo {author} {\bibfnamefont {J.}~\bibnamefont {Yao}}, \bibinfo {author} {\bibfnamefont {A.~W.}\ \bibnamefont {Sandvik}},\ and\ \bibinfo {author} {\bibfnamefont {M.~H.}\ \bibnamefont {Amin}},\ }\href {https://doi.org/10.1038/s41586-023-05867-2} {\bibfield  {journal} {\bibinfo  {journal} {Nature}\ }\textbf {\bibinfo {volume} {617}},\ \bibinfo {pages} {61} (\bibinfo {year} {2023})}\BibitemShut {NoStop}%
    \bibitem [{\citenamefont {Garcia}\ and\ \citenamefont {Chepiga}(2024)}]{garcia2024resolving}%
      \BibitemOpen
      \bibfield  {author} {\bibinfo {author} {\bibfnamefont {J.~S.}\ \bibnamefont {Garcia}}\ and\ \bibinfo {author} {\bibfnamefont {N.}~\bibnamefont {Chepiga}},\ }\href@noop {} {\bibinfo {title} {Resolving chiral transitions in rydberg arrays with quantum kibble-zurek mechanism and finite-time scaling}} (\bibinfo {year} {2024}),\ \Eprint {https://arxiv.org/abs/2403.03081} {arXiv:2403.03081 [cond-mat.str-el]} \BibitemShut {NoStop}%
    \bibitem [{\citenamefont {Dupont}\ and\ \citenamefont {Moore}(2022)}]{PhysRevB.106.L041109}%
      \BibitemOpen
      \bibfield  {author} {\bibinfo {author} {\bibfnamefont {M.}~\bibnamefont {Dupont}}\ and\ \bibinfo {author} {\bibfnamefont {J.~E.}\ \bibnamefont {Moore}},\ }\href {https://doi.org/10.1103/PhysRevB.106.L041109} {\bibfield  {journal} {\bibinfo  {journal} {Phys. Rev. B}\ }\textbf {\bibinfo {volume} {106}},\ \bibinfo {pages} {L041109} (\bibinfo {year} {2022})}\BibitemShut {NoStop}%
    \bibitem [{\citenamefont {Polkovnikov}\ and\ \citenamefont {Gritsev}(2008)}]{Polkovnikov2008natphy}%
      \BibitemOpen
      \bibfield  {author} {\bibinfo {author} {\bibfnamefont {A.}~\bibnamefont {Polkovnikov}}\ and\ \bibinfo {author} {\bibfnamefont {V.}~\bibnamefont {Gritsev}},\ }\href {https://doi.org/https://doi.org/10.1038/nphys963} {\bibfield  {journal} {\bibinfo  {journal} {Nature Physics}\ }\textbf {\bibinfo {volume} {4}},\ \bibinfo {pages} {477} (\bibinfo {year} {2008})}\BibitemShut {NoStop}%
    \bibitem [{\citenamefont {Divakaran}\ \emph {et~al.}(2008)\citenamefont {Divakaran}, \citenamefont {Dutta},\ and\ \citenamefont {Sen}}]{PhysRevB.78.144301}%
      \BibitemOpen
      \bibfield  {author} {\bibinfo {author} {\bibfnamefont {U.}~\bibnamefont {Divakaran}}, \bibinfo {author} {\bibfnamefont {A.}~\bibnamefont {Dutta}},\ and\ \bibinfo {author} {\bibfnamefont {D.}~\bibnamefont {Sen}},\ }\href {https://doi.org/10.1103/PhysRevB.78.144301} {\bibfield  {journal} {\bibinfo  {journal} {Phys. Rev. B}\ }\textbf {\bibinfo {volume} {78}},\ \bibinfo {pages} {144301} (\bibinfo {year} {2008})}\BibitemShut {NoStop}%
    \bibitem [{\citenamefont {Suzuki}\ and\ \citenamefont {Dutta}(2015)}]{PhysRevB.92.064419}%
      \BibitemOpen
      \bibfield  {author} {\bibinfo {author} {\bibfnamefont {S.}~\bibnamefont {Suzuki}}\ and\ \bibinfo {author} {\bibfnamefont {A.}~\bibnamefont {Dutta}},\ }\href {https://doi.org/10.1103/PhysRevB.92.064419} {\bibfield  {journal} {\bibinfo  {journal} {Phys. Rev. B}\ }\textbf {\bibinfo {volume} {92}},\ \bibinfo {pages} {064419} (\bibinfo {year} {2015})}\BibitemShut {NoStop}%
    \bibitem [{\citenamefont {Gross}\ and\ \citenamefont {Neveu}(1974)}]{Gross1974prd}%
      \BibitemOpen
      \bibfield  {author} {\bibinfo {author} {\bibfnamefont {D.~J.}\ \bibnamefont {Gross}}\ and\ \bibinfo {author} {\bibfnamefont {A.}~\bibnamefont {Neveu}},\ }\href {https://doi.org/10.1103/PhysRevD.10.3235} {\bibfield  {journal} {\bibinfo  {journal} {Phys. Rev. D}\ }\textbf {\bibinfo {volume} {10}},\ \bibinfo {pages} {3235} (\bibinfo {year} {1974})}\BibitemShut {NoStop}%
    \bibitem [{\citenamefont {Gracey}\ \emph {et~al.}(2016)\citenamefont {Gracey}, \citenamefont {Luthe},\ and\ \citenamefont {Schr\"oder}}]{Gracey2016prd}%
      \BibitemOpen
      \bibfield  {author} {\bibinfo {author} {\bibfnamefont {J.~A.}\ \bibnamefont {Gracey}}, \bibinfo {author} {\bibfnamefont {T.}~\bibnamefont {Luthe}},\ and\ \bibinfo {author} {\bibfnamefont {Y.}~\bibnamefont {Schr\"oder}},\ }\href {https://doi.org/10.1103/PhysRevD.94.125028} {\bibfield  {journal} {\bibinfo  {journal} {Phys. Rev. D}\ }\textbf {\bibinfo {volume} {94}},\ \bibinfo {pages} {125028} (\bibinfo {year} {2016})}\BibitemShut {NoStop}%
    \bibitem [{\citenamefont {Poland}\ \emph {et~al.}(2019)\citenamefont {Poland}, \citenamefont {Rychkov},\ and\ \citenamefont {Vichi}}]{Poland2019rmp}%
      \BibitemOpen
      \bibfield  {author} {\bibinfo {author} {\bibfnamefont {D.}~\bibnamefont {Poland}}, \bibinfo {author} {\bibfnamefont {S.}~\bibnamefont {Rychkov}},\ and\ \bibinfo {author} {\bibfnamefont {A.}~\bibnamefont {Vichi}},\ }\href {https://doi.org/10.1103/RevModPhys.91.015002} {\bibfield  {journal} {\bibinfo  {journal} {Rev. Mod. Phys.}\ }\textbf {\bibinfo {volume} {91}},\ \bibinfo {pages} {015002} (\bibinfo {year} {2019})}\BibitemShut {NoStop}%
    \bibitem [{\citenamefont {You}\ \emph {et~al.}(2018)\citenamefont {You}, \citenamefont {He}, \citenamefont {Xu},\ and\ \citenamefont {Vishwanath}}]{Youyz2018prx}%
      \BibitemOpen
      \bibfield  {author} {\bibinfo {author} {\bibfnamefont {Y.-Z.}\ \bibnamefont {You}}, \bibinfo {author} {\bibfnamefont {Y.-C.}\ \bibnamefont {He}}, \bibinfo {author} {\bibfnamefont {C.}~\bibnamefont {Xu}},\ and\ \bibinfo {author} {\bibfnamefont {A.}~\bibnamefont {Vishwanath}},\ }\href {https://doi.org/10.1103/PhysRevX.8.011026} {\bibfield  {journal} {\bibinfo  {journal} {Phys. Rev. X}\ }\textbf {\bibinfo {volume} {8}},\ \bibinfo {pages} {011026} (\bibinfo {year} {2018})}\BibitemShut {NoStop}%
    \bibitem [{\citenamefont {Castro~Neto}\ \emph {et~al.}(2009)\citenamefont {Castro~Neto}, \citenamefont {Guinea}, \citenamefont {Peres}, \citenamefont {Novoselov},\ and\ \citenamefont {Geim}}]{Geim2009rmp}%
      \BibitemOpen
      \bibfield  {author} {\bibinfo {author} {\bibfnamefont {A.~H.}\ \bibnamefont {Castro~Neto}}, \bibinfo {author} {\bibfnamefont {F.}~\bibnamefont {Guinea}}, \bibinfo {author} {\bibfnamefont {N.~M.~R.}\ \bibnamefont {Peres}}, \bibinfo {author} {\bibfnamefont {K.~S.}\ \bibnamefont {Novoselov}},\ and\ \bibinfo {author} {\bibfnamefont {A.~K.}\ \bibnamefont {Geim}},\ }\href {https://doi.org/10.1103/RevModPhys.81.109} {\bibfield  {journal} {\bibinfo  {journal} {Rev. Mod. Phys.}\ }\textbf {\bibinfo {volume} {81}},\ \bibinfo {pages} {109} (\bibinfo {year} {2009})}\BibitemShut {NoStop}%
    \bibitem [{\citenamefont {Hasan}\ and\ \citenamefont {Kane}(2010)}]{KaneReview}%
      \BibitemOpen
      \bibfield  {author} {\bibinfo {author} {\bibfnamefont {M.~Z.}\ \bibnamefont {Hasan}}\ and\ \bibinfo {author} {\bibfnamefont {C.~L.}\ \bibnamefont {Kane}},\ }\href {https://doi.org/10.1103/RevModPhys.82.3045} {\bibfield  {journal} {\bibinfo  {journal} {Rev. Mod. Phys.}\ }\textbf {\bibinfo {volume} {82}},\ \bibinfo {pages} {3045} (\bibinfo {year} {2010})}\BibitemShut {NoStop}%
    \bibitem [{\citenamefont {Qi}\ and\ \citenamefont {Zhang}(2011)}]{SCZhangReview}%
      \BibitemOpen
      \bibfield  {author} {\bibinfo {author} {\bibfnamefont {X.-L.}\ \bibnamefont {Qi}}\ and\ \bibinfo {author} {\bibfnamefont {S.-C.}\ \bibnamefont {Zhang}},\ }\href {https://doi.org/10.1103/RevModPhys.83.1057} {\bibfield  {journal} {\bibinfo  {journal} {Rev. Mod. Phys.}\ }\textbf {\bibinfo {volume} {83}},\ \bibinfo {pages} {1057} (\bibinfo {year} {2011})}\BibitemShut {NoStop}%
    \bibitem [{\citenamefont {Janssen}\ and\ \citenamefont {Herbut}(2014{\natexlab{a}})}]{Janssen2014prb}%
      \BibitemOpen
      \bibfield  {author} {\bibinfo {author} {\bibfnamefont {L.}~\bibnamefont {Janssen}}\ and\ \bibinfo {author} {\bibfnamefont {I.~F.}\ \bibnamefont {Herbut}},\ }\href {https://doi.org/10.1103/PhysRevB.89.205403} {\bibfield  {journal} {\bibinfo  {journal} {Phys. Rev. B}\ }\textbf {\bibinfo {volume} {89}},\ \bibinfo {pages} {205403} (\bibinfo {year} {2014}{\natexlab{a}})}\BibitemShut {NoStop}%
    \bibitem [{\citenamefont {Parisen~Toldin}\ \emph {et~al.}(2015{\natexlab{a}})\citenamefont {Parisen~Toldin}, \citenamefont {Hohenadler}, \citenamefont {Assaad},\ and\ \citenamefont {Herbut}}]{Parisen2015prb}%
      \BibitemOpen
      \bibfield  {author} {\bibinfo {author} {\bibfnamefont {F.}~\bibnamefont {Parisen~Toldin}}, \bibinfo {author} {\bibfnamefont {M.}~\bibnamefont {Hohenadler}}, \bibinfo {author} {\bibfnamefont {F.~F.}\ \bibnamefont {Assaad}},\ and\ \bibinfo {author} {\bibfnamefont {I.~F.}\ \bibnamefont {Herbut}},\ }\href {https://doi.org/10.1103/PhysRevB.91.165108} {\bibfield  {journal} {\bibinfo  {journal} {Phys. Rev. B}\ }\textbf {\bibinfo {volume} {91}},\ \bibinfo {pages} {165108} (\bibinfo {year} {2015}{\natexlab{a}})}\BibitemShut {NoStop}%
    \bibitem [{\citenamefont {Li}\ \emph {et~al.}(2018)\citenamefont {Li}, \citenamefont {Vaezi}, \citenamefont {Mendl},\ and\ \citenamefont {Yao}}]{Li2018ScienceAdvances}%
      \BibitemOpen
      \bibfield  {author} {\bibinfo {author} {\bibfnamefont {Z.-X.}\ \bibnamefont {Li}}, \bibinfo {author} {\bibfnamefont {A.}~\bibnamefont {Vaezi}}, \bibinfo {author} {\bibfnamefont {C.~B.}\ \bibnamefont {Mendl}},\ and\ \bibinfo {author} {\bibfnamefont {H.}~\bibnamefont {Yao}},\ }\href {https://doi.org/10.1126/sciadv.aau1463} {\bibfield  {journal} {\bibinfo  {journal} {Science Advances}\ }\textbf {\bibinfo {volume} {4}},\ \bibinfo {pages} {eaau1463} (\bibinfo {year} {2018})}\BibitemShut {NoStop}%
    \bibitem [{\citenamefont {Tabatabaei}\ \emph {et~al.}(2022)\citenamefont {Tabatabaei}, \citenamefont {Negari}, \citenamefont {Maciejko},\ and\ \citenamefont {Vaezi}}]{Vaezi2022PRL}%
      \BibitemOpen
      \bibfield  {author} {\bibinfo {author} {\bibfnamefont {S.~M.}\ \bibnamefont {Tabatabaei}}, \bibinfo {author} {\bibfnamefont {A.-R.}\ \bibnamefont {Negari}}, \bibinfo {author} {\bibfnamefont {J.}~\bibnamefont {Maciejko}},\ and\ \bibinfo {author} {\bibfnamefont {A.}~\bibnamefont {Vaezi}},\ }\href {https://doi.org/10.1103/PhysRevLett.128.225701} {\bibfield  {journal} {\bibinfo  {journal} {Phys. Rev. Lett.}\ }\textbf {\bibinfo {volume} {128}},\ \bibinfo {pages} {225701} (\bibinfo {year} {2022})}\BibitemShut {NoStop}%
    \bibitem [{\citenamefont {Otsuka}\ \emph {et~al.}(2018)\citenamefont {Otsuka}, \citenamefont {Seki}, \citenamefont {Sorella},\ and\ \citenamefont {Yunoki}}]{Sorella2018PRB}%
      \BibitemOpen
      \bibfield  {author} {\bibinfo {author} {\bibfnamefont {Y.}~\bibnamefont {Otsuka}}, \bibinfo {author} {\bibfnamefont {K.}~\bibnamefont {Seki}}, \bibinfo {author} {\bibfnamefont {S.}~\bibnamefont {Sorella}},\ and\ \bibinfo {author} {\bibfnamefont {S.}~\bibnamefont {Yunoki}},\ }\href {https://doi.org/10.1103/PhysRevB.98.035126} {\bibfield  {journal} {\bibinfo  {journal} {Phys. Rev. B}\ }\textbf {\bibinfo {volume} {98}},\ \bibinfo {pages} {035126} (\bibinfo {year} {2018})}\BibitemShut {NoStop}%
    \bibitem [{\citenamefont {Sorella}\ and\ \citenamefont {Tosatti}(1992)}]{Sorella1992}%
      \BibitemOpen
      \bibfield  {author} {\bibinfo {author} {\bibfnamefont {S.}~\bibnamefont {Sorella}}\ and\ \bibinfo {author} {\bibfnamefont {E.}~\bibnamefont {Tosatti}},\ }\href {https://doi.org/10.1209/0295-5075/19/8/007} {\bibfield  {journal} {\bibinfo  {journal} {Europhysics Letters}\ }\textbf {\bibinfo {volume} {19}},\ \bibinfo {pages} {699} (\bibinfo {year} {1992})}\BibitemShut {NoStop}%
    \bibitem [{\citenamefont {Herbut}(2006)}]{Herbut2006prl}%
      \BibitemOpen
      \bibfield  {author} {\bibinfo {author} {\bibfnamefont {I.~F.}\ \bibnamefont {Herbut}},\ }\href {https://doi.org/10.1103/PhysRevLett.97.146401} {\bibfield  {journal} {\bibinfo  {journal} {Phys. Rev. Lett.}\ }\textbf {\bibinfo {volume} {97}},\ \bibinfo {pages} {146401} (\bibinfo {year} {2006})}\BibitemShut {NoStop}%
    \bibitem [{\citenamefont {Assaad}\ and\ \citenamefont {Herbut}(2013)}]{Herbut2013prx}%
      \BibitemOpen
      \bibfield  {author} {\bibinfo {author} {\bibfnamefont {F.~F.}\ \bibnamefont {Assaad}}\ and\ \bibinfo {author} {\bibfnamefont {I.~F.}\ \bibnamefont {Herbut}},\ }\href {https://doi.org/10.1103/PhysRevX.3.031010} {\bibfield  {journal} {\bibinfo  {journal} {Phys. Rev. X}\ }\textbf {\bibinfo {volume} {3}},\ \bibinfo {pages} {031010} (\bibinfo {year} {2013})}\BibitemShut {NoStop}%
    \bibitem [{\citenamefont {Parisen~Toldin}\ \emph {et~al.}(2015{\natexlab{b}})\citenamefont {Parisen~Toldin}, \citenamefont {Hohenadler}, \citenamefont {Assaad},\ and\ \citenamefont {Herbut}}]{PhysRevB.91.165108}%
      \BibitemOpen
      \bibfield  {author} {\bibinfo {author} {\bibfnamefont {F.}~\bibnamefont {Parisen~Toldin}}, \bibinfo {author} {\bibfnamefont {M.}~\bibnamefont {Hohenadler}}, \bibinfo {author} {\bibfnamefont {F.~F.}\ \bibnamefont {Assaad}},\ and\ \bibinfo {author} {\bibfnamefont {I.~F.}\ \bibnamefont {Herbut}},\ }\href {https://doi.org/10.1103/PhysRevB.91.165108} {\bibfield  {journal} {\bibinfo  {journal} {Phys. Rev. B}\ }\textbf {\bibinfo {volume} {91}},\ \bibinfo {pages} {165108} (\bibinfo {year} {2015}{\natexlab{b}})}\BibitemShut {NoStop}%
    \bibitem [{\citenamefont {Otsuka}\ \emph {et~al.}(2016)\citenamefont {Otsuka}, \citenamefont {Yunoki},\ and\ \citenamefont {Sorella}}]{Sorella2016prx}%
      \BibitemOpen
      \bibfield  {author} {\bibinfo {author} {\bibfnamefont {Y.}~\bibnamefont {Otsuka}}, \bibinfo {author} {\bibfnamefont {S.}~\bibnamefont {Yunoki}},\ and\ \bibinfo {author} {\bibfnamefont {S.}~\bibnamefont {Sorella}},\ }\href {https://doi.org/10.1103/PhysRevX.6.011029} {\bibfield  {journal} {\bibinfo  {journal} {Phys. Rev. X}\ }\textbf {\bibinfo {volume} {6}},\ \bibinfo {pages} {011029} (\bibinfo {year} {2016})}\BibitemShut {NoStop}%
    \bibitem [{\citenamefont {Wang}\ \emph {et~al.}(2014)\citenamefont {Wang}, \citenamefont {Corboz},\ and\ \citenamefont {Troyer}}]{Wang2014NJP}%
      \BibitemOpen
      \bibfield  {author} {\bibinfo {author} {\bibfnamefont {L.}~\bibnamefont {Wang}}, \bibinfo {author} {\bibfnamefont {P.}~\bibnamefont {Corboz}},\ and\ \bibinfo {author} {\bibfnamefont {M.}~\bibnamefont {Troyer}},\ }\href {https://doi.org/10.1088/1367-2630/16/10/103008} {\bibfield  {journal} {\bibinfo  {journal} {New Journal of Physics}\ }\textbf {\bibinfo {volume} {16}},\ \bibinfo {pages} {103008} (\bibinfo {year} {2014})}\BibitemShut {NoStop}%
    \bibitem [{\citenamefont {Li}\ \emph {et~al.}(2015{\natexlab{a}})\citenamefont {Li}, \citenamefont {Jiang},\ and\ \citenamefont {Yao}}]{Li2015NJP}%
      \BibitemOpen
      \bibfield  {author} {\bibinfo {author} {\bibfnamefont {Z.-X.}\ \bibnamefont {Li}}, \bibinfo {author} {\bibfnamefont {Y.-F.}\ \bibnamefont {Jiang}},\ and\ \bibinfo {author} {\bibfnamefont {H.}~\bibnamefont {Yao}},\ }\href {https://doi.org/10.1088/1367-2630/17/8/085003} {\bibfield  {journal} {\bibinfo  {journal} {New Journal of Physics}\ }\textbf {\bibinfo {volume} {17}},\ \bibinfo {pages} {085003} (\bibinfo {year} {2015}{\natexlab{a}})}\BibitemShut {NoStop}%
    \bibitem [{\citenamefont {Li}\ \emph {et~al.}(2017)\citenamefont {Li}, \citenamefont {Jiang}, \citenamefont {Jian},\ and\ \citenamefont {Yao}}]{Yao2017nc}%
      \BibitemOpen
      \bibfield  {author} {\bibinfo {author} {\bibfnamefont {Z.-X.}\ \bibnamefont {Li}}, \bibinfo {author} {\bibfnamefont {Y.-F.}\ \bibnamefont {Jiang}}, \bibinfo {author} {\bibfnamefont {S.-K.}\ \bibnamefont {Jian}},\ and\ \bibinfo {author} {\bibfnamefont {H.}~\bibnamefont {Yao}},\ }\href {https://doi.org/10.1038/s41467-017-00167-6} {\bibfield  {journal} {\bibinfo  {journal} {Nature Communications}\ }\textbf {\bibinfo {volume} {8}},\ \bibinfo {pages} {314} (\bibinfo {year} {2017})}\BibitemShut {NoStop}%
    \bibitem [{\citenamefont {Dutta}\ \emph {et~al.}()\citenamefont {Dutta}, \citenamefont {Singh},\ and\ \citenamefont {Divakaran}}]{dutta_quenching_2010}%
      \BibitemOpen
      \bibfield  {author} {\bibinfo {author} {\bibfnamefont {A.}~\bibnamefont {Dutta}}, \bibinfo {author} {\bibfnamefont {R.~R.~P.}\ \bibnamefont {Singh}},\ and\ \bibinfo {author} {\bibfnamefont {U.}~\bibnamefont {Divakaran}},\ }\href {https://doi.org/10.1209/0295-5075/89/67001} {\bibfield  {journal} {\bibinfo  {journal} {Europhysics Letters}\ }\textbf {\bibinfo {volume} {89}},\ \bibinfo {pages} {67001}}\BibitemShut {NoStop}%
    \bibitem [{\citenamefont {Sun}\ \emph {et~al.}(2022)\citenamefont {Sun}, \citenamefont {Deng},\ and\ \citenamefont {Li}}]{PhysRevB.106.134203}%
      \BibitemOpen
      \bibfield  {author} {\bibinfo {author} {\bibfnamefont {Z.}~\bibnamefont {Sun}}, \bibinfo {author} {\bibfnamefont {M.}~\bibnamefont {Deng}},\ and\ \bibinfo {author} {\bibfnamefont {F.}~\bibnamefont {Li}},\ }\href {https://doi.org/10.1103/PhysRevB.106.134203} {\bibfield  {journal} {\bibinfo  {journal} {Phys. Rev. B}\ }\textbf {\bibinfo {volume} {106}},\ \bibinfo {pages} {134203} (\bibinfo {year} {2022})}\BibitemShut {NoStop}%
    \bibitem [{\citenamefont {Deng}\ \emph {et~al.}(2025)\citenamefont {Deng}, \citenamefont {Sun},\ and\ \citenamefont {Li}}]{PhysRevLett.134.010409}%
      \BibitemOpen
      \bibfield  {author} {\bibinfo {author} {\bibfnamefont {M.}~\bibnamefont {Deng}}, \bibinfo {author} {\bibfnamefont {Z.}~\bibnamefont {Sun}},\ and\ \bibinfo {author} {\bibfnamefont {F.}~\bibnamefont {Li}},\ }\href {https://doi.org/10.1103/PhysRevLett.134.010409} {\bibfield  {journal} {\bibinfo  {journal} {Phys. Rev. Lett.}\ }\textbf {\bibinfo {volume} {134}},\ \bibinfo {pages} {010409} (\bibinfo {year} {2025})}\BibitemShut {NoStop}%
    \bibitem [{\citenamefont {Assaad}\ and\ \citenamefont {Evertz}(2008)}]{AssaadReview}%
      \BibitemOpen
      \bibfield  {author} {\bibinfo {author} {\bibfnamefont {F.}~\bibnamefont {Assaad}}\ and\ \bibinfo {author} {\bibfnamefont {H.}~\bibnamefont {Evertz}},\ }\bibinfo {title} {World-line and determinantal quantum monte carlo methods for spins, phonons and electrons},\ in\ \href {https://doi.org/10.1007/978-3-540-74686-7_10} {\emph {\bibinfo {booktitle} {Computational Many-Particle Physics}}},\ \bibinfo {editor} {edited by\ \bibinfo {editor} {\bibfnamefont {H.}~\bibnamefont {Fehske}}, \bibinfo {editor} {\bibfnamefont {R.}~\bibnamefont {Schneider}},\ and\ \bibinfo {editor} {\bibfnamefont {A.}~\bibnamefont {Wei{\ss}e}}}\ (\bibinfo  {publisher} {Springer Berlin Heidelberg},\ \bibinfo {address} {Berlin, Heidelberg},\ \bibinfo {year} {2008})\ pp.\ \bibinfo {pages} {277--356}\BibitemShut {NoStop}%
    \bibitem [{\citenamefont {Li}\ and\ \citenamefont {Yao}(2019)}]{Li2019Review}%
      \BibitemOpen
      \bibfield  {author} {\bibinfo {author} {\bibfnamefont {Z.-X.}\ \bibnamefont {Li}}\ and\ \bibinfo {author} {\bibfnamefont {H.}~\bibnamefont {Yao}},\ }\href {https://doi.org/10.1146/annurev-conmatphys-033117-054307} {\bibfield  {journal} {\bibinfo  {journal} {Annual Review of Condensed Matter Physics}\ }\textbf {\bibinfo {volume} {10}},\ \bibinfo {pages} {337} (\bibinfo {year} {2019})},\ \Eprint {https://arxiv.org/abs/https://doi.org/10.1146/annurev-conmatphys-033117-054307} {https://doi.org/10.1146/annurev-conmatphys-033117-054307} \BibitemShut {NoStop}%
    \bibitem [{\citenamefont {Schmitt}\ \emph {et~al.}(2022)\citenamefont {Schmitt}, \citenamefont {Rams}, \citenamefont {Dziarmaga}, \citenamefont {Heyl},\ and\ \citenamefont {Zurek}}]{Dziarmaga2022sciadv}%
      \BibitemOpen
      \bibfield  {author} {\bibinfo {author} {\bibfnamefont {M.}~\bibnamefont {Schmitt}}, \bibinfo {author} {\bibfnamefont {M.~M.}\ \bibnamefont {Rams}}, \bibinfo {author} {\bibfnamefont {J.}~\bibnamefont {Dziarmaga}}, \bibinfo {author} {\bibfnamefont {M.}~\bibnamefont {Heyl}},\ and\ \bibinfo {author} {\bibfnamefont {W.~H.}\ \bibnamefont {Zurek}},\ }\href {https://doi.org/10.1126/sciadv.abl6850} {\bibfield  {journal} {\bibinfo  {journal} {Science Advances}\ }\textbf {\bibinfo {volume} {8}},\ \bibinfo {pages} {eabl6850} (\bibinfo {year} {2022})},\ \Eprint {https://arxiv.org/abs/https://www.science.org/doi/pdf/10.1126/sciadv.abl6850} {https://www.science.org/doi/pdf/10.1126/sciadv.abl6850} \BibitemShut {NoStop}%
    \bibitem [{\citenamefont {De~Grandi}\ \emph {et~al.}(2011)\citenamefont {De~Grandi}, \citenamefont {Polkovnikov},\ and\ \citenamefont {Sandvik}}]{Polkovnikov2011prb}%
      \BibitemOpen
      \bibfield  {author} {\bibinfo {author} {\bibfnamefont {C.}~\bibnamefont {De~Grandi}}, \bibinfo {author} {\bibfnamefont {A.}~\bibnamefont {Polkovnikov}},\ and\ \bibinfo {author} {\bibfnamefont {A.~W.}\ \bibnamefont {Sandvik}},\ }\href {https://doi.org/10.1103/PhysRevB.84.224303} {\bibfield  {journal} {\bibinfo  {journal} {Phys. Rev. B}\ }\textbf {\bibinfo {volume} {84}},\ \bibinfo {pages} {224303} (\bibinfo {year} {2011})}\BibitemShut {NoStop}%
    \bibitem [{\citenamefont {Hesselmann}\ and\ \citenamefont {Wessel}(2016)}]{Hesselmann2016prb}%
      \BibitemOpen
      \bibfield  {author} {\bibinfo {author} {\bibfnamefont {S.}~\bibnamefont {Hesselmann}}\ and\ \bibinfo {author} {\bibfnamefont {S.}~\bibnamefont {Wessel}},\ }\href {https://doi.org/10.1103/PhysRevB.93.155157} {\bibfield  {journal} {\bibinfo  {journal} {Phys. Rev. B}\ }\textbf {\bibinfo {volume} {93}},\ \bibinfo {pages} {155157} (\bibinfo {year} {2016})}\BibitemShut {NoStop}%
    \bibitem [{\citenamefont {Li}\ \emph {et~al.}(2015{\natexlab{b}})\citenamefont {Li}, \citenamefont {Jiang},\ and\ \citenamefont {Yao}}]{Li2015PRB}%
      \BibitemOpen
      \bibfield  {author} {\bibinfo {author} {\bibfnamefont {Z.-X.}\ \bibnamefont {Li}}, \bibinfo {author} {\bibfnamefont {Y.-F.}\ \bibnamefont {Jiang}},\ and\ \bibinfo {author} {\bibfnamefont {H.}~\bibnamefont {Yao}},\ }\href {https://doi.org/10.1103/PhysRevB.91.241117} {\bibfield  {journal} {\bibinfo  {journal} {Phys. Rev. B}\ }\textbf {\bibinfo {volume} {91}},\ \bibinfo {pages} {241117} (\bibinfo {year} {2015}{\natexlab{b}})}\BibitemShut {NoStop}%
    \bibitem [{\citenamefont {Li}\ \emph {et~al.}(2016)\citenamefont {Li}, \citenamefont {Jiang},\ and\ \citenamefont {Yao}}]{Li2016PRL}%
      \BibitemOpen
      \bibfield  {author} {\bibinfo {author} {\bibfnamefont {Z.-X.}\ \bibnamefont {Li}}, \bibinfo {author} {\bibfnamefont {Y.-F.}\ \bibnamefont {Jiang}},\ and\ \bibinfo {author} {\bibfnamefont {H.}~\bibnamefont {Yao}},\ }\href {https://doi.org/10.1103/PhysRevLett.117.267002} {\bibfield  {journal} {\bibinfo  {journal} {Phys. Rev. Lett.}\ }\textbf {\bibinfo {volume} {117}},\ \bibinfo {pages} {267002} (\bibinfo {year} {2016})}\BibitemShut {NoStop}%
    \bibitem [{\citenamefont {Wei}\ \emph {et~al.}(2016)\citenamefont {Wei}, \citenamefont {Wu}, \citenamefont {Li}, \citenamefont {Zhang},\ and\ \citenamefont {Xiang}}]{Xiang2016PRL}%
      \BibitemOpen
      \bibfield  {author} {\bibinfo {author} {\bibfnamefont {Z.~C.}\ \bibnamefont {Wei}}, \bibinfo {author} {\bibfnamefont {C.}~\bibnamefont {Wu}}, \bibinfo {author} {\bibfnamefont {Y.}~\bibnamefont {Li}}, \bibinfo {author} {\bibfnamefont {S.}~\bibnamefont {Zhang}},\ and\ \bibinfo {author} {\bibfnamefont {T.}~\bibnamefont {Xiang}},\ }\href {https://doi.org/10.1103/PhysRevLett.116.250601} {\bibfield  {journal} {\bibinfo  {journal} {Phys. Rev. Lett.}\ }\textbf {\bibinfo {volume} {116}},\ \bibinfo {pages} {250601} (\bibinfo {year} {2016})}\BibitemShut {NoStop}%
    \bibitem [{\citenamefont {Herbut}\ \emph {et~al.}(2009)\citenamefont {Herbut}, \citenamefont {Juri\ifmmode \check{c}\else \v{c}\fi{}i\ifmmode~\acute{c}\else \'{c}\fi{}},\ and\ \citenamefont {Vafek}}]{Herbut2009prb}%
      \BibitemOpen
      \bibfield  {author} {\bibinfo {author} {\bibfnamefont {I.~F.}\ \bibnamefont {Herbut}}, \bibinfo {author} {\bibfnamefont {V.}~\bibnamefont {Juri\ifmmode \check{c}\else \v{c}\fi{}i\ifmmode~\acute{c}\else \'{c}\fi{}}},\ and\ \bibinfo {author} {\bibfnamefont {O.}~\bibnamefont {Vafek}},\ }\href {https://doi.org/10.1103/PhysRevB.80.075432} {\bibfield  {journal} {\bibinfo  {journal} {Phys. Rev. B}\ }\textbf {\bibinfo {volume} {80}},\ \bibinfo {pages} {075432} (\bibinfo {year} {2009})}\BibitemShut {NoStop}%
    \bibitem [{\citenamefont {Tang}\ \emph {et~al.}(2018)\citenamefont {Tang}, \citenamefont {Leaw}, \citenamefont {Rodrigues}, \citenamefont {Herbut}, \citenamefont {Sengupta}, \citenamefont {Assaad},\ and\ \citenamefont {Adam}}]{doi:10.1126/science.aao2934}%
      \BibitemOpen
      \bibfield  {author} {\bibinfo {author} {\bibfnamefont {H.-K.}\ \bibnamefont {Tang}}, \bibinfo {author} {\bibfnamefont {J.~N.}\ \bibnamefont {Leaw}}, \bibinfo {author} {\bibfnamefont {J.~N.~B.}\ \bibnamefont {Rodrigues}}, \bibinfo {author} {\bibfnamefont {I.~F.}\ \bibnamefont {Herbut}}, \bibinfo {author} {\bibfnamefont {P.}~\bibnamefont {Sengupta}}, \bibinfo {author} {\bibfnamefont {F.~F.}\ \bibnamefont {Assaad}},\ and\ \bibinfo {author} {\bibfnamefont {S.}~\bibnamefont {Adam}},\ }\href {https://doi.org/10.1126/science.aao2934} {\bibfield  {journal} {\bibinfo  {journal} {Science}\ }\textbf {\bibinfo {volume} {361}},\ \bibinfo {pages} {570} (\bibinfo {year} {2018})},\ \Eprint {https://arxiv.org/abs/https://www.science.org/doi/pdf/10.1126/science.aao2934} {https://www.science.org/doi/pdf/10.1126/science.aao2934} \BibitemShut {NoStop}%
    \bibitem [{\citenamefont {J{\"o}rdens}\ \emph {et~al.}(2008)\citenamefont {J{\"o}rdens}, \citenamefont {Strohmaier}, \citenamefont {G{\"u}nter}, \citenamefont {Moritz},\ and\ \citenamefont {Esslinger}}]{Esslinger2008Nature}%
      \BibitemOpen
      \bibfield  {author} {\bibinfo {author} {\bibfnamefont {R.}~\bibnamefont {J{\"o}rdens}}, \bibinfo {author} {\bibfnamefont {N.}~\bibnamefont {Strohmaier}}, \bibinfo {author} {\bibfnamefont {K.}~\bibnamefont {G{\"u}nter}}, \bibinfo {author} {\bibfnamefont {H.}~\bibnamefont {Moritz}},\ and\ \bibinfo {author} {\bibfnamefont {T.}~\bibnamefont {Esslinger}},\ }\href {https://doi.org/10.1038/nature07244} {\bibfield  {journal} {\bibinfo  {journal} {Nature}\ }\textbf {\bibinfo {volume} {455}},\ \bibinfo {pages} {204} (\bibinfo {year} {2008})}\BibitemShut {NoStop}%
    \bibitem [{\citenamefont {Mazurenko}\ \emph {et~al.}(2017)\citenamefont {Mazurenko}, \citenamefont {Chiu}, \citenamefont {Ji}, \citenamefont {Parsons}, \citenamefont {Kan{\'a}sz-Nagy}, \citenamefont {Schmidt}, \citenamefont {Grusdt}, \citenamefont {Demler}, \citenamefont {Greif},\ and\ \citenamefont {Greiner}}]{Greiner2017Nature}%
      \BibitemOpen
      \bibfield  {author} {\bibinfo {author} {\bibfnamefont {A.}~\bibnamefont {Mazurenko}}, \bibinfo {author} {\bibfnamefont {C.~S.}\ \bibnamefont {Chiu}}, \bibinfo {author} {\bibfnamefont {G.}~\bibnamefont {Ji}}, \bibinfo {author} {\bibfnamefont {M.~F.}\ \bibnamefont {Parsons}}, \bibinfo {author} {\bibfnamefont {M.}~\bibnamefont {Kan{\'a}sz-Nagy}}, \bibinfo {author} {\bibfnamefont {R.}~\bibnamefont {Schmidt}}, \bibinfo {author} {\bibfnamefont {F.}~\bibnamefont {Grusdt}}, \bibinfo {author} {\bibfnamefont {E.}~\bibnamefont {Demler}}, \bibinfo {author} {\bibfnamefont {D.}~\bibnamefont {Greif}},\ and\ \bibinfo {author} {\bibfnamefont {M.}~\bibnamefont {Greiner}},\ }\href {https://doi.org/10.1038/nature22362} {\bibfield  {journal} {\bibinfo  {journal} {Nature}\ }\textbf {\bibinfo {volume} {545}},\ \bibinfo {pages} {462} (\bibinfo {year} {2017})}\BibitemShut {NoStop}%
    \bibitem [{\citenamefont {Venu}\ \emph {et~al.}(2023)\citenamefont {Venu}, \citenamefont {Xu}, \citenamefont {Mamaev}, \citenamefont {Corapi}, \citenamefont {Bilitewski}, \citenamefont {D'Incao}, \citenamefont {Fujiwara}, \citenamefont {Rey},\ and\ \citenamefont {Thywissen}}]{Venu2023Nature}%
      \BibitemOpen
      \bibfield  {author} {\bibinfo {author} {\bibfnamefont {V.}~\bibnamefont {Venu}}, \bibinfo {author} {\bibfnamefont {P.}~\bibnamefont {Xu}}, \bibinfo {author} {\bibfnamefont {M.}~\bibnamefont {Mamaev}}, \bibinfo {author} {\bibfnamefont {F.}~\bibnamefont {Corapi}}, \bibinfo {author} {\bibfnamefont {T.}~\bibnamefont {Bilitewski}}, \bibinfo {author} {\bibfnamefont {J.~P.}\ \bibnamefont {D'Incao}}, \bibinfo {author} {\bibfnamefont {C.~J.}\ \bibnamefont {Fujiwara}}, \bibinfo {author} {\bibfnamefont {A.~M.}\ \bibnamefont {Rey}},\ and\ \bibinfo {author} {\bibfnamefont {J.~H.}\ \bibnamefont {Thywissen}},\ }\href {https://doi.org/10.1038/s41586-022-05405-6} {\bibfield  {journal} {\bibinfo  {journal} {Nature}\ }\textbf {\bibinfo {volume} {613}},\ \bibinfo {pages} {262} (\bibinfo {year} {2023})}\BibitemShut {NoStop}%
    \bibitem [{\citenamefont {Yin}\ \emph {et~al.}(2016)\citenamefont {Yin}, \citenamefont {Lo},\ and\ \citenamefont {Chen}}]{PhysRevB.94.064302}%
      \BibitemOpen
      \bibfield  {author} {\bibinfo {author} {\bibfnamefont {S.}~\bibnamefont {Yin}}, \bibinfo {author} {\bibfnamefont {C.-Y.}\ \bibnamefont {Lo}},\ and\ \bibinfo {author} {\bibfnamefont {P.}~\bibnamefont {Chen}},\ }\href {https://doi.org/10.1103/PhysRevB.94.064302} {\bibfield  {journal} {\bibinfo  {journal} {Phys. Rev. B}\ }\textbf {\bibinfo {volume} {94}},\ \bibinfo {pages} {064302} (\bibinfo {year} {2016})}\BibitemShut {NoStop}%
    \bibitem [{\citenamefont {Grandi}\ \emph {et~al.}(2013)\citenamefont {Grandi}, \citenamefont {Polkovnikov},\ and\ \citenamefont {Sandvik}}]{De_Grandi_2013}%
      \BibitemOpen
      \bibfield  {author} {\bibinfo {author} {\bibfnamefont {C.~D.}\ \bibnamefont {Grandi}}, \bibinfo {author} {\bibfnamefont {A.}~\bibnamefont {Polkovnikov}},\ and\ \bibinfo {author} {\bibfnamefont {A.~W.}\ \bibnamefont {Sandvik}},\ }\href {https://doi.org/10.1088/0953-8984/25/40/404216} {\bibfield  {journal} {\bibinfo  {journal} {Journal of Physics: Condensed Matter}\ }\textbf {\bibinfo {volume} {25}},\ \bibinfo {pages} {404216} (\bibinfo {year} {2013})}\BibitemShut {NoStop}%
    \bibitem [{\citenamefont {Kaul}(2015)}]{PhysRevLett.115.157202}%
      \BibitemOpen
      \bibfield  {author} {\bibinfo {author} {\bibfnamefont {R.~K.}\ \bibnamefont {Kaul}},\ }\href {https://doi.org/10.1103/PhysRevLett.115.157202} {\bibfield  {journal} {\bibinfo  {journal} {Phys. Rev. Lett.}\ }\textbf {\bibinfo {volume} {115}},\ \bibinfo {pages} {157202} (\bibinfo {year} {2015})}\BibitemShut {NoStop}%
    \bibitem [{\citenamefont {Rosenstein}\ \emph {et~al.}(1993)\citenamefont {Rosenstein}, \citenamefont {{Hoi-Lai Yu}},\ and\ \citenamefont {Kovner}}]{ROSENSTEIN1993381}%
      \BibitemOpen
      \bibfield  {author} {\bibinfo {author} {\bibfnamefont {B.}~\bibnamefont {Rosenstein}}, \bibinfo {author} {\bibnamefont {{Hoi-Lai Yu}}},\ and\ \bibinfo {author} {\bibfnamefont {A.}~\bibnamefont {Kovner}},\ }\href {https://doi.org/https://doi.org/10.1016/0370-2693(93)91253-J} {\bibfield  {journal} {\bibinfo  {journal} {Physics Letters B}\ }\textbf {\bibinfo {volume} {314}},\ \bibinfo {pages} {381} (\bibinfo {year} {1993})}\BibitemShut {NoStop}%
    \bibitem [{\citenamefont {Janssen}\ and\ \citenamefont {Herbut}(2014{\natexlab{b}})}]{PhysRevB.89.205403}%
      \BibitemOpen
      \bibfield  {author} {\bibinfo {author} {\bibfnamefont {L.}~\bibnamefont {Janssen}}\ and\ \bibinfo {author} {\bibfnamefont {I.~F.}\ \bibnamefont {Herbut}},\ }\href {https://doi.org/10.1103/PhysRevB.89.205403} {\bibfield  {journal} {\bibinfo  {journal} {Phys. Rev. B}\ }\textbf {\bibinfo {volume} {89}},\ \bibinfo {pages} {205403} (\bibinfo {year} {2014}{\natexlab{b}})}\BibitemShut {NoStop}%
    \bibitem [{\citenamefont {Rosa}\ \emph {et~al.}(2001)\citenamefont {Rosa}, \citenamefont {Vitale},\ and\ \citenamefont {Wetterich}}]{rosa2001}%
      \BibitemOpen
      \bibfield  {author} {\bibinfo {author} {\bibfnamefont {L.}~\bibnamefont {Rosa}}, \bibinfo {author} {\bibfnamefont {P.}~\bibnamefont {Vitale}},\ and\ \bibinfo {author} {\bibfnamefont {C.}~\bibnamefont {Wetterich}},\ }\href {https://doi.org/10.1103/PhysRevLett.86.958} {\bibfield  {journal} {\bibinfo  {journal} {Phys. Rev. Lett.}\ }\textbf {\bibinfo {volume} {86}},\ \bibinfo {pages} {958} (\bibinfo {year} {2001})}\BibitemShut {NoStop}%
    \end{thebibliography}
\end{document}